\pdfoutput=1
\documentclass[twocolumn,amsmath,aps,floats,amssymb, floatfix, superscriptaddress, nofootinbib, secnumarabic]{revtex4}
\usepackage{graphicx}
\usepackage{wasysym}
\usepackage{bmpsize}
\usepackage{amsfonts}
\usepackage{subfigure}
\usepackage{hyperref}
\usepackage{amsmath}
\usepackage{color}        
\usepackage{ulem}
\usepackage{lipsum}
\usepackage{gensymb}
\usepackage{longtable}
\usepackage{comment}
\usepackage{multirow}
\usepackage{units}
\usepackage{float}
\usepackage[export]{adjustbox}

\newcommand{\sfig}[2]{
	\includegraphics[width=#2]{#1}
}

\newcommand{\Sfig}[2]{
    \begin{figure}[thbp]
    \sfig{#1}{1.\columnwidth}
    \caption{{\small #2}}
    \label{fig:#1}
    \end{figure}
}

\newcommand{\Sfigbig}[2]{
	\begin{figure*}[thbp]
		\sfig{#1}{2.\columnwidth}
		\caption{{\small #2}}
		\label{fig:#1}
	\end{figure*}
}

\newcommand{\Rf}[1]{Fig.~(\ref{fig:#1})}
\newcommand{\rf}[1]{\ref{fig:#1}}
\newcommand\losen{line-of-sight extrapolation noise}
\newcommand{\be}{\begin{equation}}
\newcommand{\ee}{\end{equation}}
\def\ba#1\ea{\begin{align}\begin{split}#1\end{split}\end{align}}

\def\bea{\begin{eqnarray}}
\def\eea{\end{eqnarray}}
\newcommand{\ec}[1]{Eq.~(\ref{eq:#1})}
\newcommand{\Ec}[1]{(\ref{eq:#1})}
\newcommand{\eql}[1]{\label{eq:#1}}

\begin{document}

\title{Line-of-Sight Extrapolation Noise in Dust Polarization}         
\author{Jason Poh}
\affiliation{Department of Astronomy \& Astrophysics, University of Chicago, Chicago, IL 60637}
\affiliation{Kavli Institute for Cosmological Physics, Enrico Fermi Institute, University of Chicago, Chicago, IL 60637}
\author{Scott Dodelson}        
\affiliation{Fermi National Accelerator Laboratory, Batavia, IL 60510-0500}
\affiliation{Kavli Institute for Cosmological Physics, Enrico Fermi Institute, University of Chicago, Chicago, IL 60637}
\affiliation{Department of Astronomy \& Astrophysics, University of Chicago, Chicago, IL 60637}

\begin{abstract}
\noindent
The B-modes of polarization at frequencies ranging from 50-1000 GHz are produced by Galactic dust, lensing of primordial E-modes in the cosmic microwave background (CMB) by intervening large scale structure, and possibly by primordial B-modes in the CMB imprinted by gravitational waves produced during inflation. The conventional method used to separate the dust component of the signal is to assume that the signal at high frequencies (e.g., 350 GHz) is due solely to dust and then extrapolate the signal down to a lower frequency (e.g., 150 GHz) using the measured scaling of the polarized dust signal amplitude with frequency. For typical Galactic thermal dust temperatures of $\sim$20K, these frequencies are not fully in the Rayleigh-Jeans limit. Therefore, deviations in the dust cloud temperatures from cloud to cloud will lead to different scaling factors for clouds of different temperatures. Hence, when multiple clouds of different temperatures and polarization angles contribute to the integrated line-of-sight polarization signal, the relative contribution of individual clouds to the integrated signal can change between frequencies. This can cause the integrated signal to be decorrelated in both amplitude and direction when extrapolating in frequency. Here we carry out a Monte Carlo analysis on the impact of this \textit{line-of-sight extrapolation noise}, enabling us to quantify its effect. Using results from the Planck experiment, we find that this effect is small, more than an order of magnitude smaller than the current uncertainties. However, \losen\ may be a significant source of uncertainty in future low-noise primordial B-mode experiments. Scaling from Planck results, we find that accounting for this uncertainty becomes potentially important when experiments are sensitive to primordial B-mode signals with amplitude $r\lesssim$ 0.0015 . 
  \end{abstract}

\maketitle

\section{Introduction}

Primordial B-modes in the cosmic microwave background (CMB) are an important signature of inflation\cite{Kamionkowski1996,Kamionkowski1997,Seljak1996,Seljak1996a}. In the inflationary paradigm, the very early universe underwent a period of exponential expansion, generating gravitational waves that were eventually imprinted on the polarization of the CMB at last scattering on degree angular scales. A detection of this primordial B-mode signal would be strong evidence for inflation, and the strength of the detected signal would aid in constraining inflation models. 

In practice, however, the detection of primordial B-modes is complicated by the fact that any primordial signal is likely contaminated by foreground sources, and separating out the contributions from each of these sources will require great care. At degree angular scales and frequencies above 100 GHz, the primary foreground contribution comes from linearly polarized emission from asymmetric dust grains that are aligned with the Galactic magnetic field \cite{PlanckCollaboration2015a,PlanckCollaboration2014,Draine2003,Lazarian2007}.

Empirically, the polarized dust spectral emission distribution is typically fitted as a modified blackbody in the frequency range targeted by CMB experiments ($100-1000$ GHz). Hence, a common technique used to remove polarized dust emission in CMB experiments is to use polarization maps measured at higher frequencies (e.g., $\sim$350 GHz), where the polarized dust emission signal dominates, to infer dust polarization properties at lower frequencies targeted by CMB experiments (e.g., $\sim$150 GHz). Using the modified blackbody parametrization, the spectral energy distribution (SED) of polarized dust emission scales with frequency as a product of the blackbody spectral radiance and an empirically fitted spectral index, $I(\nu) \propto B(\nu, T)\nu^{\beta}$, with the polarization angle remaining unchanged between frequencies. The extrapolated dust map at 150 GHz can then be used to separate the polarized dust emission component from other contributions. 

This separation technique has been used in recent analyses of B-modes in CMB experiments such as Planck and BICEP2. In those analyses, the polarized dust SED was determined to be well-fitted by a single mean dust temperature and spectral index (e.g. see \cite{Ade2015c,PlanckCollaboration2014}).  However, for the frequencies of interest in CMB experiments, the thermal SED for typical dust temperatures of $\sim$20 K is not fully in the Rayleigh-Jeans regime. Therefore, the polarized SED can have a significant dependence on the dust temperature (\Rf{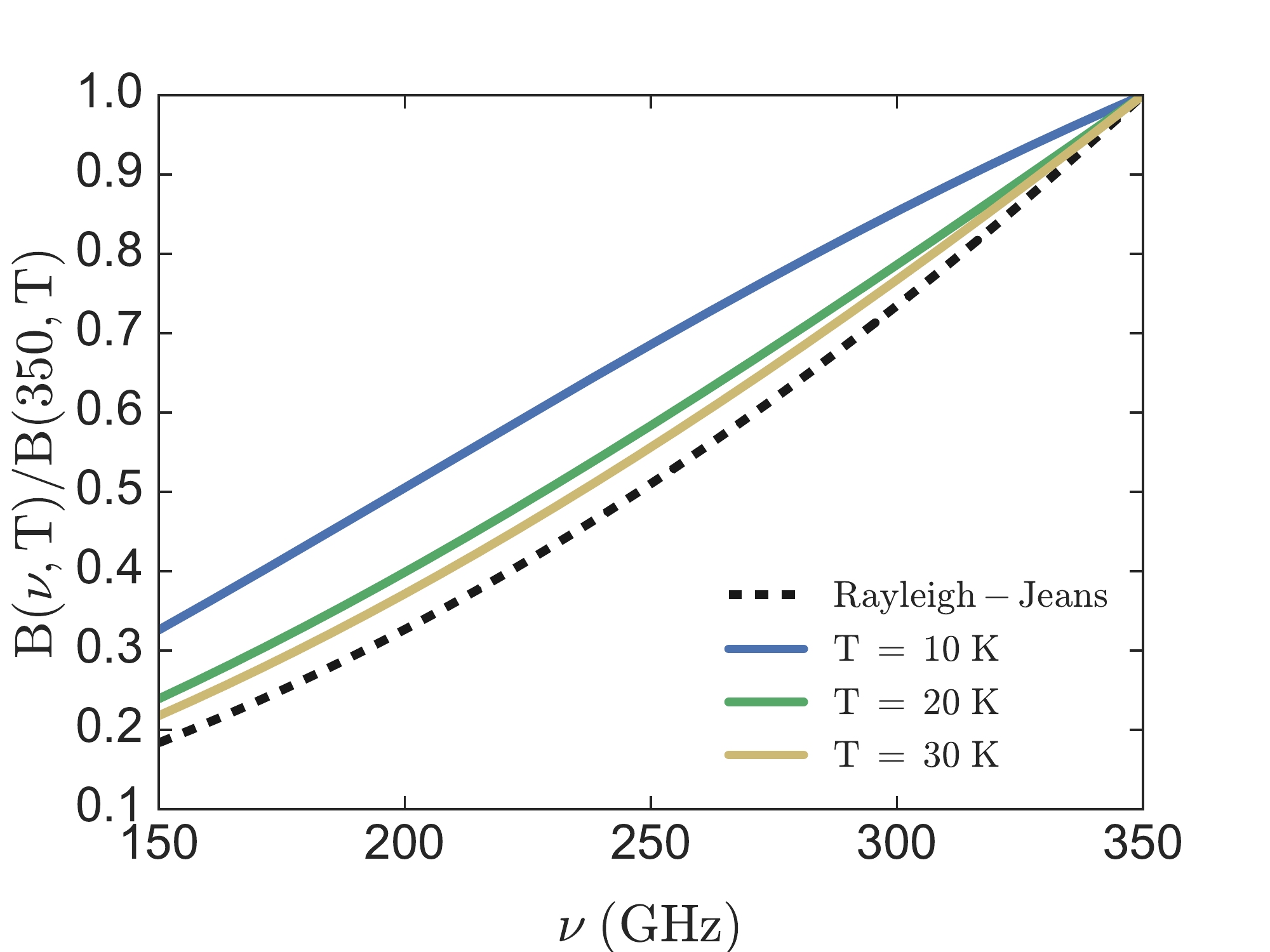}). Since the temperature distribution of Galactic dust varies from cloud to cloud (we define a dust cloud as any dusty region with a single characteristic temperature, column density and polarization angle), the SEDs of individual clouds scale differently with frequency. 

If there are multiple clouds along a line-of-sight, the dust polarization signal observed by a CMB experiment combines the polarized emissions from each contributing cloud along the line-of-sight. The frequency extrapolation is carried out on this integrated polarized signal. If the contributing dust clouds along a line-of-sight have different temperatures, then the relative contribution to the integrated dust polarization signal from each dust cloud changes with frequency. In principle, this can lead to large deviations from the assumed power law scaling factor in the dust polarization signal, especially if the polarization angles of the contributing dust clouds are severely misaligned. In addition, the polarization angle and polarization fraction of the integrated polarization signal can be significantly decorrelated between frequencies. This effect, which produces spatial variations in the integrated polarized dust SED, was previously explored using a two-cloud model \cite{Tassis2015}. 

The goal of this paper is to understand how much this effect, which we call the {\it line-of-sight extrapolation noise}, impacts the estimates of low frequency polarization. In \S\ref{theory}, we review the basic idea of why extrapolating the dust signal between different frequencies can fail. Then in \S\ref{method}, we describe our methodology for characterizing \losen\ in the polarized dust emission observables. In \S\ref{results}, we present estimates of \losen\ in various observables using empirically motivated dust distribution models and compare these with Planck results in \S\ref{Planckcomparison} to estimate how important this effect is compared to other sources of uncertainty. We close with some remarks on implications for current and future experiments in \S\ref{future}. 

\section{\label{theory}Dust Model and Assumptions}

The polarization of thermal dust emission can be described in terms of Stokes $I$, $Q$ and $U$ parameters. In the optically thin regime, the specific intensity $I$ of a dust cloud at a frequency $\nu_a$ is empirically well-fitted by the modified blackbody parameterization
\ba \eql{Inu}
	I(\nu_a) &=B(\nu_a,T)\,\kappa(\nu_a)\,N_{d} \\
	& \propto B(\nu_a, T)\,\nu_a^{\beta}
\ea
where 
\be
	B(\nu_a, T) = \frac{2h\nu_a^3}{c^2} \frac{1}{\exp({\frac{h\nu_a}{k_B T}}) - 1}
\ee
is the blackbody spectral radiance; $N_{d}$ is the dust column density; and $T$ is the temperature of the dust cloud. The dust opacity $\kappa$ is usually described as a power law \cite{Hildebrand1983,Abergel2014}
\be \eql{kappa}
\kappa(\nu_a)= \kappa_0 \left(\frac{\nu_a}{\nu_0}\right)^{\beta}
\ee
where $\kappa_0$ is the dust opacity at some reference frequency $\nu_0$ and $\beta$ is the spectral index. The Stokes $Q$ and $U$ parameters are
\ba \eql{singlestokes}
Q(\nu_a) &= p I(\nu_a) \cos(2\alpha)\\ 
U(\nu_a) &= p I(\nu_a) \sin(2\alpha) \\
\ea
where $p$ is the polarization fraction of the cloud (here, we make the simplifying assumption that it is independent of frequency); $\alpha$ is the angle of polarization with respect to a reference axis; and $I(\nu_a)$ is the specific intensity at $\nu_a$, given by \ec{Inu}. 

To extrapolate the Stokes parameters $S \in \{I,Q,U\}$ of the polarized dust emission from that observed at frequency $\nu_a$ to a lower frequency $\nu_b$, we use equations \Ec{Inu}-\Ec{singlestokes} to obtain the estimator
\ba 
\hat S(\nu_b) &= S(\nu_a) \frac{B(\nu_b, T)}{B(\nu_a, T)}   \left(\frac{\nu_b}{\nu_a}\right)^{\beta}  \eql{est} .
\ea
\ec{est} is the explicit parameterization of the estimator used to extrapolate the polarized dust SED in the Planck collaboration's analysis of thermal dust polarization data (see e.g. \cite{Abergel2011,Abergel2014,Ade2015,Ade2015b}). \Rf{RJ.pdf} shows that the ratio of the radiance at two frequencies is temperature dependent.

If there are multiple clouds along one line-of-sight, the total integrated Stokes parameter along the line-of-sight is the sum of individual contributions from each cloud. For example, the integrated Stokes $Q$ parameter is
\ba \eql{sumQ}
Q({\nu_a}) &= \sum_i Q_{a,i}  = \sum_i I_i(\nu_a)\,p_i \cos(2\alpha_i) \\
\ea
 where the index $i$ labels individual clouds. If we assume that the dust opacity law in \ec{kappa} is universal (i.e. the spectral index $\beta$ is the same for all dust clouds), we can remove the opacity 
term from the sum and write \ec{sumQ} as 
\ba \eql{los}
Q({\nu_a}) = \kappa(\nu_a)\,\sum_i\, B(\nu_a,T_i)\,N_{d,i}\, p_i\,\cos(2\alpha_i) .
\ea
Therefore, the ratio of $Q$ at two frequencies is
\ba \eql{losratio} 
\frac{Q(\nu_b)}{Q(\nu_a)}  &= \frac{ \kappa(\nu_b)}{ \kappa(\nu_a)} \frac{\sum_i \, B(\nu_b,T_i)\,N_{d,i}\,p_i\cos(2\alpha_i)}{\sum_i\,B(\nu_a,T_i)\, N_{d,i}\,p_i\cos(2\alpha_i)} \\
&= \left(\frac{\nu_b}{\nu_a}\right)^{\beta} \frac{\sum_i\, B(\nu_b,T_i)\,N_{d,i}\,p_i\cos(2\alpha_i)}{\sum_i\, B(\nu_a,T_i)\, N_{d,i}\,p_i\cos(2\alpha_i)} 
\ea
Following the same treatment, the ratio of the Stokes $I$ and $U$ parameters at two frequencies are
\begin{eqnarray} 
\frac{I(\nu_b)}{I(\nu_a)}  &=& \left(\frac{\nu_b}{\nu_a}\right)^{\beta} \frac{\sum_i\, B(\nu_b,T_i)\,N_{d,i}}{\sum_i\, B(\nu_a,T_i)\, N_{d,i}} 
\eql{losratioIU} \\
\frac{U(\nu_b)}{U(\nu_a)}  &=& \left(\frac{\nu_b}{\nu_a}\right)^{\beta} \frac{\sum_i\, B(\nu_b,T_i)\,N_{d,i}\,p_i\sin(2\alpha_i)}{\sum_i\, B(\nu_a,T_i)\, N_{d,i}\,p_i\sin(2\alpha_i)} 
\end{eqnarray}

\Sfig{RJ.pdf}{Ratio of the blackbody spectral radiance (as a function of frequency)  to the blackbody spectral radiance at 350 GHz for three different dust temperatures. The Rayleigh-Jeans law is plotted for comparison. In the limit of large temperature ($T\gg 30K$), the plotted ratio is independent of temperature, approaching the Rayleigh-Jeans scaling of $\nu^2$. But at typical diffuse dust cloud temperatures of $\sim$20K, the scaling of the blackbody spectral radiance function with frequency has a significant temperature dependence. For example, note the difference between this ratio at $\nu=150$ GHz when the temperature is 10K as opposed to 30K.}	

Generally, equations \Ec{losratio}-\Ec{losratioIU}, which represent the true frequency scaling relation when there are multiple cloud contributions along a line-of-sight,  do not reduce to \ec{est} except in three cases: (1) there is only 1 cloud along the line-of-sight,  (2) every cloud along the line-of-sight has the same temperature, or (3) the polarized dust SED is deep in the Rayleigh-Jeans regime. The first two cases are physically unrealistic, since we expect there to be multiple clouds along a line-of-sight simply from Copernican arguments, as empirical dust emission maps from Planck show marked angular variation in dust temperature (see e.g.~\cite{Ade2015}). If the polarized dust emission were deep in the Rayleigh-Jeans regime, then the Planck function $B(\nu_,T_i)  \rightarrow 2\nu^2k_BT/c^2$. In that case, $B(\nu,T)$ is a pure power law in frequency, and the frequency-dependent part of $B(\nu,T_i)$ can be factored out of the sum in both numerator and denominator, and the remaining terms in the sum cancel out. In this scenario, both equations \Ec{est} and \Ec{losratio}--\Ec{losratioIU} reduce to the same power law in frequency with exponent $\gamma=\beta +2$. Therefore, \ec{est} would be a perfect estimator. However, the frequencies targeted by CMB experiments and typical dust cloud temperatures do not fall within the Rayleigh-Jeans regime (see \Rf{RJ.pdf}). 

The estimator in \ec{est} therefore deviates non-linearly from the true scaling factor given by equations \Ec{losratio}--\Ec{losratioIU}, resulting in some degree of extrapolation error when the estimator is used, which we refer to as \losen. The polarization fraction and polarization angle of the line-of-sight integrated polarized dust signal can therefore be significantly decorrelated between different frequencies. The net polarization fraction $p$ and polarization angle $\alpha$ at frequency $\nu_a$ are related to the net Stokes parameters along a line-of-sight by 
\ba \eql{pol}
p(\nu_a) = \frac{\sqrt{Q(\nu_a)^2 + U(\nu_a)^2}}{I(\nu_a)} \\
\alpha(\nu_a) = \frac{1}{2}\tan^{-1}\left(\frac{U(\nu_a)}{Q(\nu_a)}\right)
\ea
If \ec{est} is a perfect estimator, then the frequency-dependent factors $B(\nu,T)\,\nu^{\beta}$ in the Stokes $Q$, $U$, and $I$ parameters exactly cancel out in the numerator and denominator in the equations for $p(\nu_a)$ and $\alpha(\nu_a)$, leaving the polarization fraction and angle unchanged between frequencies (i.e. $p(\nu_a)=p(\nu_b)=p$ and  $\alpha(\nu_a) = \alpha(\nu_b) = \alpha$). However, in actuality, the frequency dependence of the integrated Stokes parameters (Eqs. \Ec{losratio}-\Ec{losratioIU}) in \ec{pol} does not trivially vanish. Therefore, the polarization fraction and angle can differ between two frequencies. 

It is worth emphasizing that this \losen\ is a systematic error that affects all CMB experiments and \textit{cannot} simply be reduced by virtue of better instrument resolution or sensitivity alone. Therefore, it is imperative that the extent of this effect is characterized, and its effect on the accuracy of the inferred polarized dust foreground emissions at frequencies targeted by CMB experiments is well-understood. 

The degree of \losen\ depends non-trivially on the cloud properties, such as the number of contributing clouds and temperature of the clouds along the line-of-sight, which makes characterizing and subtracting this effect challenging. A previous study using a two-cloud model demonstrated that the \losen\ can be potentially large in scenarios where the polarization angles of the contributing clouds along a line-of-sight are significantly misaligned with respect to each other \cite{Tassis2015}. In this model, if the relative contribution to the integrated polarization signal of the clouds changes between frequencies, then the polarization signal of the first cloud may dominate at one frequency, while the polarization signal of the second cloud may dominate at a different frequency, leading to decorrelated polarization properties between the two frequencies if the clouds are severely misaligned with respect to each other. However, the true statistical significance of this source of uncertainty is not yet well-understood, especially in a more general model where there are multiple contributions along each line-of-sight. Therefore, a more robust analysis of the statistical significance of this source of uncertainty for many lines-of-sight is required. 

In the remainder of this paper, we describe a first step towards the statistical characterization of this \losen, using a single population of dust clouds which is assumed to be well-described by a single universal modified blackbody SED. This simplifying assumption was implied in our analytic expressions \ec{losratio}-\ec{losratioIU}, where we assumed a universal dust opacity law for every dust grain along the line-of-sight. In reality, dust grain populations are heterogeneous, with varying dust compositions, grain sizes, and orientation with respect to local radiation/magnetic field geometries in the Galaxy (e.g. \cite{Compiegne2010,Jones2013,Hildebrand1999,Draine2009,Draine2012,Draine2013}), all of which can result in different dust SEDs. The integrated thermal dust SED for these multi-component dust populations is therefore likely to have more complex dependencies on, e.g. $T$, $p$, $N_d$, $\alpha$, than we have described in \ec{losratio}-\ec{losratioIU}. This is likely to affect our overall characterization of the \losen. However, due to the current lack of observational constraints on the large-scale distribution and statistical properties of these multifarious dust grain populations, we leave these considerations to a future study.

\section{\label{method}Methodology} 

To quantify the statistical significance of \losen, we perform a Monte Carlo analysis as follows: first, we simulate a mock sky map corresponding to a region that may be targeted by a CMB instrument. Every pixel on the map represents one line-of-sight, and each line-of-sight contains some number of contributing clouds. The number of clouds along each line-of-sight is allowed to vary, as are the temperatures, cloud column densities, polarization fractions and polarization angles of each cloud. The integrated $I$, $Q$ and $U$ Stokes parameters from polarized dust emission are then calculated for every pixel by summing the Stokes parameters of each cloud using \ec{losratio} to obtain the integrated dust polarization signal. The polarization fraction and angle of the integrated signal in each pixel are then calculated using \ec{pol}. 

The above process is carried out at 150 GHz and 350 GHz. These two frequencies were chosen to coincide with the frequency used in the BICEP2 experiment and the frequency of the dust polarization map used by the Planck experiment to estimate the dust polarization signal in the BICEP2 field \cite{Ade2015d,PlanckCollaboration2014,Ade2015c}. These simulated maps represent the \textit{true} thermal polarized dust signal at those two frequencies. We then calculate an \textit{inferred} temperature for each line-of-sight by fitting for the integrated $I_{True}(150)$ and  $I_{True}(350)$ signal with the estimator \ec{est}, assuming some fiducial spectral index $\beta$. For this study, we use the Planck value of $\beta = 1.59$ \cite{Ade2015b}. We then take the true polarized dust map at 350 GHz and scale the amplitude of the signal in each pixel according to the estimator \ec{est} to the target frequency of 150 GHz using the inferred temperature and the fiducial spectral index $\beta = 1.59$. The resulting map represents the \textit{predicted} thermal dust polarization map at 150 GHz from extrapolation. The pixel-by-pixel deviation of the predicted polarization properties from the true polarization properties at 150 GHz are then calculated. The statistical properties of the \losen\ are then evaluated for the simulated sky map.

We quantify \losen\ using three observables: 
\begin{enumerate}
	\item $Q_{\mathrm{True}}/Q_{\mathrm{predicted}}$
	\item $p_{\mathrm{True}}/p_{\mathrm{predicted}}$
	\item $\alpha_{\mathrm{True}} - \alpha_{\mathrm{predicted}}$ 
\end{enumerate}
If the extrapolation is perfect, then we expect (1) $Q_{\mathrm{True}}/Q_{\mathrm{predicted}} = 1$, (2) $p_{\mathrm{True}}/p_{\mathrm{predicted}} = 1$, and (3) $|\alpha_{\mathrm{True}} - \alpha_{\mathrm{predicted}}| = 0$ for each pixel\footnote{Alternatively, subtracting 1 from each value of $Q_{\mathrm{True}}/Q_{\mathrm{predicted}}$ (likewise for $p$) converts it into a measure of the fractional difference $ Q_{\mathrm{True}}/Q_{\mathrm{predicted}}-1 = (Q_{\mathrm{True}}-Q_{\mathrm{predicted}}) /Q_{\mathrm{predicted}} $.}. Therefore, we quantify \losen\ using the statistical scatter of these parameters from their expected values for all the pixels in a given sky region. The Stokes $U$ parameter is also used to determine the polarization fraction and angle in our model, but will not be included as an quantity of interest in this analysis. The reason for this omission is because we assume the dust clouds have a uniform random distribution of polarization angles from an arbitrary reference axis in our models (see discussion \S{\ref{dustpol}}). As such, there is no preferred directionality in the polarization angles, and hence no meaningful physical or statistical distinction between the Stokes $Q$ and $U$ parameters. 

Following Ref.~\cite{Tassis2015}, we define the ratio of cloud intensities along each line of sight as
\ba \eql{rpara}
r_i(\nu_a) &= \frac{I_i(\nu_a)}{I_0(\nu_a)} \\
&= \frac{B(\nu_a,T_i)N_{d,i}}{B(\nu_a,T_0)N_{d,0}}
\ea
where $i$ refers to the $i_{th}$ cloud along the line-of-sight and $0$ is some arbitrary reference cloud along the line-of-sight. This definition allows us to parameterize \ec{los} (and similarly for the Stokes $U$ parameter) in terms of the specific intensity of the reference cloud, $I_0$:

\ba \eql{newpara}
Q({\nu_a}) &= I_0(\nu_a) \sum_i r_i (\nu_a) p_i\, \cos(2\alpha_i) \\
\ea
This parametrization has an advantage over \ec{los} in that it depends on the dimensionless ratio of column densities $N_{d,i}/N_{d,0}$ instead of the actual column densities. Since the main source of information about the dust column densities come from dust extinction data, the ratio of dust extinctions can serve as a direct proxy of the dust column density ratios without requiring any normalization. This simplifies the number of input parameters required to determine the Stokes parameters to the following 5 parameters:
\begin{enumerate}
	\item Number of distinct clouds
	\item Column density ratio, $N_{d,i}/N_{d,0}$ 
	\item Temperature, $T_i$
	\item Polarization fraction, $p_i$
	\item Polarization angle, $\alpha_i$
\end{enumerate}

For our Monte Carlo analysis, we use two different 3D maps of the distribution of dust clouds, which we discuss in further detail in \S{\ref{dustdist}}. For both sky maps, we emulate the analysis done by the Planck Collaboration \cite{Ade2015b} and use the HEALpix software \cite{Gorski2005} to analyze the sky polarization maps at a resolution of $N_{\mathrm{side}} = 128$, corresponding to an angular resolution of 27'.5. We focus on a circular patch of 30$\degree$ radius  containing 13284 pixels centered on the Galactic pole. Each pixel represents a line-of-sight, and each contribution along the line-of-sight is assigned a temperature, polarization fraction, column density ratio and polarization angle, drawn from an empirically-motivated distribution. The polarized dust emission properties can then be calculated. Details of the distribution of these parameters are discussed in \S{\ref{dustpol}}. Table \ref{fiducialtable} summarizes the models and the input parameters used in this analysis. 

\begin{table*}[htbp] 
	\begin{ruledtabular}
		\begin{tabular}{ c  l  l  }
			\textbf{Parameter}  & \textbf{Poisson distribution model (model 1)} & \textbf{3D Pan-STARR1 reddening map (model 2)} \\
			\hline \\
			\parbox{2cm}{ Number of contributing clouds} & Poisson distribution with a mean of 9 clouds per kpc & 13 logarithmic distance bins out to 1 kpc \\
			$N_{d,i}/N_{d,0}$ & \parbox{8.0cm}{\flushleft Gaussian distribution in $\log_{10}(N_{d,i}/N_{d,0})$ with mean 0 and standard deviation 0.42} & Fixed by line-of-sight reddening profiles  \\ 
			$T$ &  \parbox{8cm}{\flushleft{Gaussian with mean $T_{mean} = 19.56$ and standard deviation $\sigma_{T} = 3.19$}} & \parbox{8cm}{\flushleft{Gaussian with mean $T_{mean} = 19.64$ and standard deviation $\sigma_{T} = 3.45$}}\\ 
			$p$ & \parbox{8cm}{\flushleft{Gaussian with mean $p_{mean} = 0.146$ and standard deviation $\sigma_{p} = 0.03$}} & \parbox{8cm}{\flushleft{Gaussian with mean $p_{mean} = 0.157$ and standard deviation $\sigma_{p} = 0.03$} }  \\
			$\alpha$ & \parbox{8cm}{\flushleft{Uniform random distribution}} & \parbox{8cm}{\flushleft{Uniform random distribution }}  \\
		\end{tabular}
	\end{ruledtabular}
	\caption{\label{fiducialtable}Summary of fiducial dust distribution model parameters used in two different dust distribution models to characterize \losen. The distributions for the temperature and polarization fraction were obtained by fitting the model to reproduce the integrated temperature distribution. Additional details of the fitting procedure and how the parameters for each model were derived are discussed in detail in \S{\ref{dustdist}}. Specific values of the fitted mean and standard deviations of the Gaussian distribution for $T$ and $p$ in various dust distribution models are presented in Table \ref{TPtable}.}
\end{table*}

\subsection{Dust Cloud Models: Number and Density} \label{dustdist}

In this section, we introduce two different dust cloud distribution models, focusing on the number of clouds along each line of sight and their column densities. The first and simpler model assumes that there are a discrete number of contributing clouds along every line-of-sight, sampled from a Poisson distribution. The second model uses distance and reddening data from Pan-STARRS 1 and 2MASS photometry to infer dust column densities at different distance bins \cite{Green2015}. We discuss the details of the two models below. 

\subsubsection{\label{poissondist}Poisson Cloud Distribution (Model 1)}

\Sfig{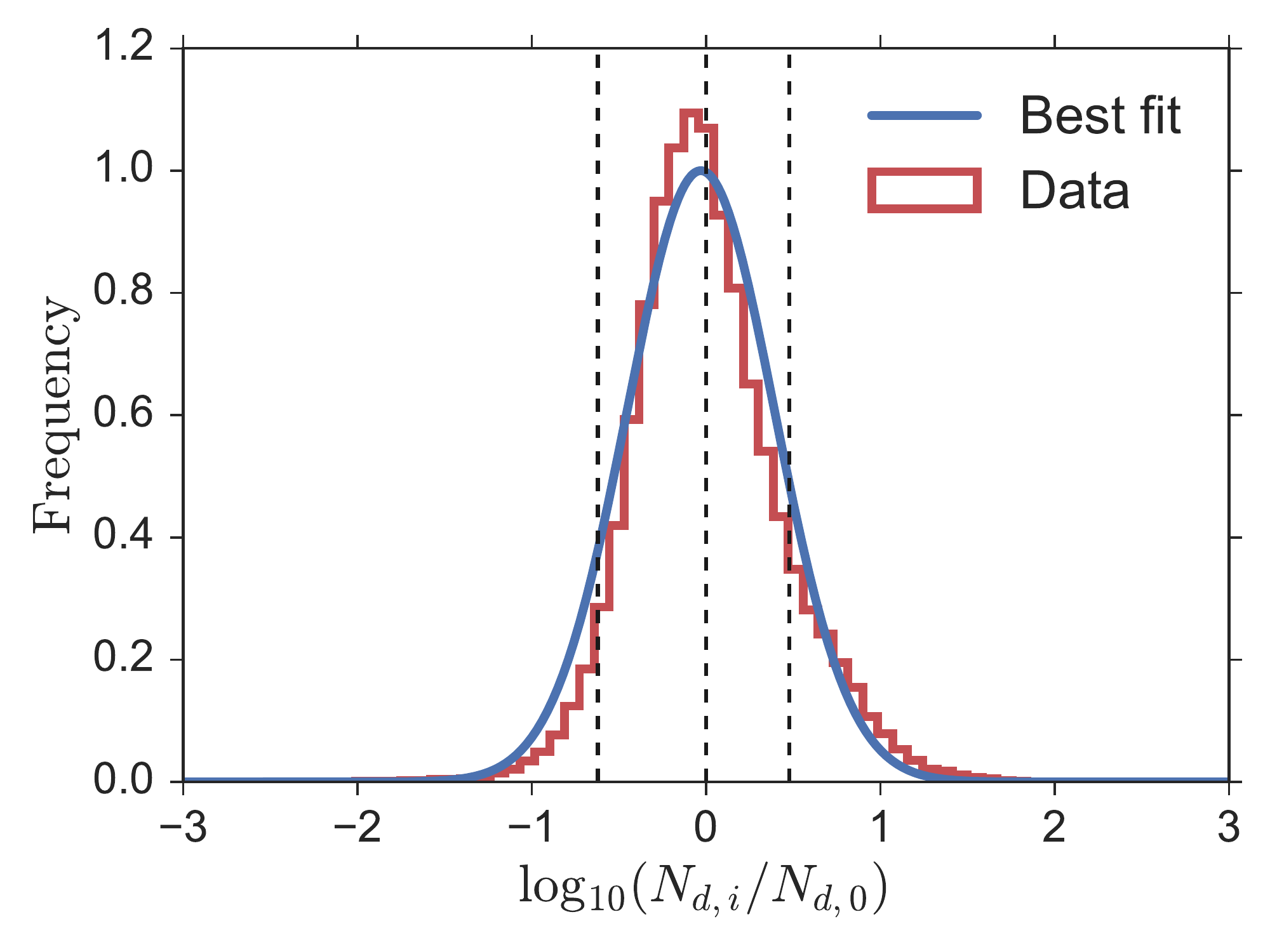}{Histogram of the quantity $\log_{10}(N_{d,i}/N_{d,0})$ from Pan-STARRS 1 and 2MASS photometry data, where $N_{d,0}$ is the median cloud column density defined for each line of sight. The best fit Gaussian, with a mean of 0 and standard deviation of 0.42, is overplotted in blue. Dashed vertical lines indicate values of $\log_{10}(N_{d,i}/N_{d,0})$ for the three-cloud model by Vergely et al. \cite{Vergely1998}. Further discussion on the data analysis is provided in \S{\ref{poissondist}}.}

This model assumes that each line-of-sight contains a discrete number of dust clouds drawn from a Poisson distribution. Previous statistical studies of the extinction in the solar neighborhood suggest that stellar extinction observations are best fit by three kinds of cloud with different extinctions: weak extinction clouds with $E(B-V) = 0.012$, medium extinction clouds with $E(B-V) = 0.05$ and dark clouds with $E(B-V) > 0.1$ \cite{Vergely1998}. The total cloud distribution in this model follows a Poisson distribution with 9 clouds per kpc. Since tomographic studies of the Milky Way suggest a characteristic disk scale height   $\lesssim1$kpc (e.g. \cite{Juric2008}), we assume there are no clouds outside of 1 kpc.

The dust column density can be inferred from the dust extinction, with the ratios of cloud extinctions serving as a proxy for the cloud column density ratios. In the above study, the three cloud types corresponded to three different characteristic extinction values. However, instead of using these values, we use the best-fit distribution from the higher resolution Pan-STARRS 1 and 2MASS photometry data \cite{Green2015}. 

From Pan-STARRS 1 data, we take the cumulative reddening data in 13 distance bins out to 1 kpc and convert it to non-cumulative reddening in each distance bin. We then set any reddening value of E(B-V) $<0.001$ to 0 as those values are likely to be spurious. For convention, we set the reference cloud to be the cloud with median extinction along each line-of-sight, and then calculate the logarithm of the ratio $\log_{10}(N_{d,i}/N_{d,0})$ for every cloud, where $N_{d,0}$ is the reference cloud along each line-of-sight. This is repeated for every line-of-sight in the 30$\degree$ radius sky patch centered on the North Galactic Pole, which is representative of regions targeted by CMB experiments. 

 The resulting distribution is best fitted by a Gaussian with mean 0 and standard deviation 0.42 (see \Rf{ratio.pdf}). The best-fit values for the three-cloud model are also indicated by the vertical dashed lines, where we use the medium extinction cloud as the reference cloud and set the characteristic extinction of the dark clouds to $E(B-V) \approx 0.15$, following typical best fit values in \cite{Vergely1998}. We find that Pan-STARRS 1 extinction data agrees with the three-cloud model, as demonstrated by the fact that the best-fit values for the three-cloud model fall within the distribution described by the Pan-STARRS 1 extinction data. Therefore, in our model, we sample the column density ratio for each cloud from the Pan-STARRS 1 distribution.

\subsubsection{\label{PSdist}Pan-STARRS 1 Stellar Photometry (Model 2)}

\begin{figure}[htbp]
	\centering
	\subfigure[$ $ \label{fig:EBVmap.pdf}  Orthographic projection of dust reddening map at 1 kpc, centered on the North Galactic Pole.]{\includegraphics[scale=0.55, valign = c]{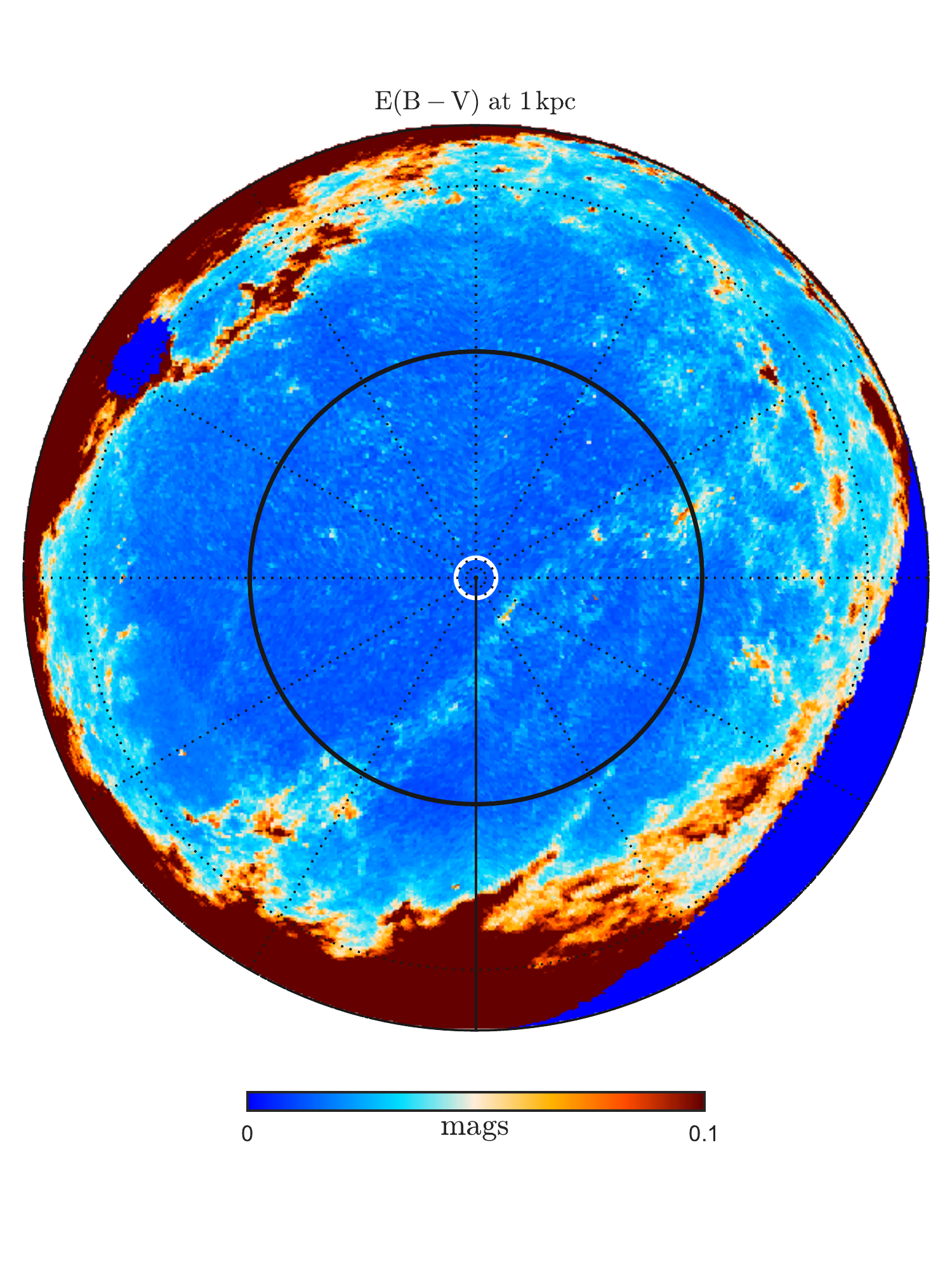}}
	\subfigure[$ $ \label{fig:EBV.pdf}  Reddening out to 1 kpc for 100 lines-of-sight near the North Galactic Pole. ]{\includegraphics[scale=0.4, valign = c]{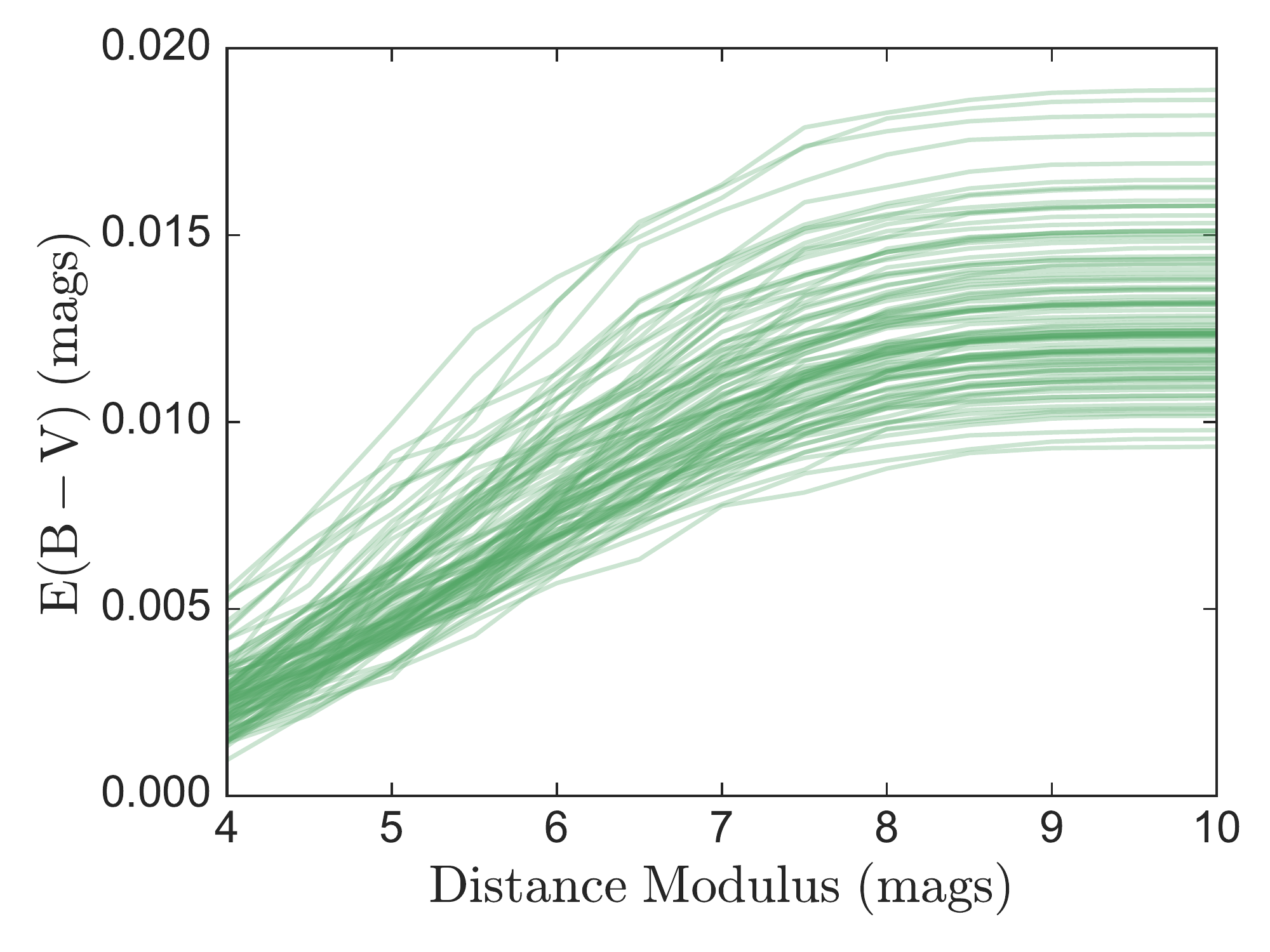}}
	\caption{ Top: Integrated reddening map at 1 kpc from Pan-STARRS 1 photometry. The black contour line indicates the $30\degree$ region used in our analysis. Bottom: Reddening as a function of distance for 100 lines-of-sight near the North Galactic Pole out to 1 kpc. The white contour line in (a) indicates where these sight lines are located.}
\end{figure}

We used 3D dust reddening-distance maps from Pan-STARRS 1 and 2MASS photometry \cite{Green2015} to infer the dust distribution in a 30$\degree$ radius region centered at the Northern Galactic Pole (\Rf{EBVmap.pdf}), as regions near the Galactic Poles are most likely to be targeted by CMB experiments. Reddening data is available for 31 logarithmic distance bins along each line-of-sight out to a distance modulus of $\mu = 19.0$ or $\sim63$ kpc. Each distance bin is treated as a discrete contribution to the polarized emission along the line-of-sight. 

As with the Poisson model, we use the reddening information as a proxy for the dust column density. The increase in reddening between distance bins is taken to be proportional to the dust column density in that bin. Likewise, we set any reddening value of E(B-V) $<0.001$ to 0 as those values are likely to be spurious. However, unlike the Poisson model, we use the reddening data directly instead of drawing randomly from a fitted distribution model. Because the dust reddening maps have varying angular resolution, we first upsample the map to the maximum HEALPix resolution on the map, $N_{\mathrm{side}} = 2048$, and then downsample the map to the target resolution of $N_{\mathrm{side}} = 128$. The reddening in each downsampled pixel is obtained by taking the average reddening of all the upsampled pixels within each downsampled pixel, except for pixels for which there is no reddening data. 

For our fiducial model, we used the best-fit reddening-distance data for 13 distance bins out to a distance modulus of $\mu = 10.0$ or 1 kpc. This is a fairly conservative cut, as the reddening does not increase after 1 kpc for the vast majority of the sight lines in the $30\degree$ region near the North Galactic Pole that we consider in this analysis. For example, \Rf{EBV.pdf} plots integrated reddening for 100 lines-of-sight in the region marked by the white contour line in \Rf{EBVmap.pdf}. 

\subsection{\label{dustpol} $T$, $p$ and $\alpha$ Distributions}

To fully determine the polarization properties, the cloud temperatures, polarization fractions and polarization angles have to be specified for every dust cloud in both 3D dust maps described above. Presently, 3D maps of the dust polarization and temperature properties do not exist. However, the Planck Collaboration has produced several all-sky studies of the 2D line-of-sight integrated polarized thermal emission from dust foregrounds (e.g. \cite{Abergel2011,Abergel2014,Ade2015,Ade2015b}). Using the same modified blackbody parametrization described in \S{\ref*{theory}}, the Planck collaboration has produced statistical distributions of the inferred dust temperature $T$ for the entire sky and polarization fraction $p$ for a large region of the sky \cite{Abergel2014,Ade2015}. 

The inferred all-sky dust temperature distribution from integrated line-of-sight data is relatively uniform, with a small overall dispersion. The distribution profile at 5' angular resolution is approximately Gaussian, with a mean dust temperature and standard deviation of 19.7 $\pm$ 1.4 K for the whole sky (see Fig. 16 from \cite{Abergel2014}). The polarization fraction distribution is considerably more complex, since it is more strongly correlated with Galactic magnetic field structure and hence exhibits a larger degree of spatial and angular correlations. For the present study, we make the simplifying assumption that the polarization fraction is uncorrelated between nearby lines-of-sight and between clouds along a line-of-sight. Additionally, we assume the distribution follows a truncated Gaussian distribution with mean polarization fraction and standard deviation of 0.06 $\pm$ 0.03, where these fiducial values are approximate fits to estimates from Planck data \cite{Ade2015}. While this may be an oversimplification of the true observed distribution of polarization fractions, we find that the choice of polarization fraction distribution itself only weakly affects the overall \losen, and therefore is not an important factor in this study (see \S{\ref{senseinput}).

In reality, the true 3D cloud temperature and polarization fraction distributions likely have a larger dispersion compared to the line-of-sight integrated distributions, since line-of-sight integration effectively smooths out variations in cloud properties along the line-of-sight. Using the above distributions of  cloud temperatures and polarization fractions, we infer the true 3D distributions of these quantities for a specified dust cloud distribution recursively. We vary the initial 3D distributions and calculate the integrated Stokes parameters for every line-of-sight at 150 GHz and 350 GHz. For temperature, we use \ec{est} to fit the observed line-of-sight $T$ and polarization fraction $p$ for the Stokes parameters at these two frequencies for the fiducial spectral index $\beta = 1.59$. We then fit a Gaussian distribution to the resulting distribution and perform a $\chi^2$ minimization to get the initial 3D distribution to produce the observed line-of-sight temperature distribution of $T$ = 19.7 $\pm$ 1.4 K. The 3D polarization fraction distribution is inferred in a similar manner, but using only the generated 350 GHz Stokes parameters. We use the integrated Stokes parameter to calculate the integrated polarization fraction and fit the initial conditions so as to reproduce the model distribution of $p$ = 0.06 $\pm$ 0.03.  Specific values of the fitted mean and standard deviations of the Gaussian distribution for $T$ and $p$ in various dust distribution models are presented in Table \ref{TPtable} in the appendix. 

Finally, we make the simplifying assumption that the polarization angles are uncorrelated along the line-of-sight and sample the polarization angle of each cloud from a uniform random distribution. In reality, the polarization angle traces Galactic structure and magnetic field lines, so we also expect some correlation in the polarization angles of dust clouds in regions where there are prominent Galactic structures or magnetic fields. Even though CMB experiments target high Galactic latitude regions to avoid these structures, studies of Galactic dust at high latitudes using data from Planck as well as experiments like the Galactic Arecibo L-Band Feed Array HI (GALFA-HI) suggest that some degree of structural coherence in polarization angles exists even in those high latitudes regions \cite{PlanckCollaboration2016,Clark2015}. Since large \losen\ is most likely when there is significant misalignment of the polarization angles of the contributing clouds along a line-of-sight, we expect this assumption to result in an overestimation of the \losen. This possible bias is studied in more detail in \S{\ref{senseinput}. 
 
However, it is unclear how significant the structural coherence in polarization angle is in the context of our model, which considers dust contributions out to a distance of 1 kpc. Statistical studies of the polarization angle dispersion by the Planck collaboration show that the polarization angle dispersion increases by about 10$\degree$ over an angular scale of $2.5\degree$ (i.e. on average, the polarization angle direction changes by about $10\degree$ over an angular distance of $2.5\degree$) From the Pan-STARRS 1 reddening data, most of the increase in reddening near the Galactic pole occurs on distance scales of a few hundred parsecs (e.g. see \Rf{EBV.pdf}). If we make the conservative estimate that the dust polarization map measured by Planck comes from Galactic dust at 500 pc, an angular scale of 2.5$\degree$ corresponds to a physical scale length of about 20 pc, which is the size of the smallest distance bin in the Pan-STARRS 1 dust maps. If the polarization angle direction changes by about $10\degree$ over 20 pc, we do not expect dust clouds to be significantly correlated in polarization angles if they are separated by distances larger than about 100 pc.  

More generally, a limitation of our model is that since we draw values of temperature, polarization fraction and polarization angle for each cloud along the line-of-sight in this 3D model from an observationally constrained distribution without taking into account spatial information, we do not capture the effects of coherent structures in the Galactic dust that may result in correlations in $T$, $p$ and $\alpha$ between dust clouds. A more physically representative dust model might encode information about the spatial coherence of these parameters (for example, in the form of a 2-point correlation function). Generally, we expect coherence in these parameters to reduce the extent of the \losen. However, we omit these considerations in the present study. 

In \S{\ref{senseinput}}, we analyze the dependence of the \losen\ on the input distributions of cloud temperatures, polarization fractions, and polarization angle. We find that \losen\ has a strong dependence on the temperature distribution and the polarization angle dispersion, while it has a much weaker dependence on the choice of polarization fraction distribution. 

\section{\label{results}Results}

\subsection{Fiducial Models}\label{resfid}

\begin{table*} 
	\begin{ruledtabular}
		\begin{tabular}{c c c c c c c}
			Parameter & \multicolumn{3}{c}{Model 1} & \multicolumn{3}{c}{Model 2} \\ 
			\cline{2-4} \cline{5-7}
			& Median & 68\% C.L. & 95\% C.L. & Median & 68\% C.L. & 95\% C.L. \\
			\hline
			$Q_{\mathrm{True}}/Q_{\mathrm{predicted}}$ & $1.00$& $(+0.06, -0.06)$ & $(+0.50,-0.47)$ & $0.99$ &$ (+0.07, -0.07)$ & $(+0.56,-0.55)$ \\
			$p_{\mathrm{True}}/p_{\mathrm{predicted}}$ & $1.00$ & $(+0.03, -0.03)$ & $(+0.12,-0.11)$ & $0.99$& $(+0.03, -0.04)$ & $(+0.14, -0.13)$ \\
			$\alpha_{\mathrm{True}} - \alpha_{\mathrm{predicted}}\ (\degree)$ & $0.00$& $(+0.85,-0.85)$ & $(+3.50,-3.55)$ & $0.01$ & $(+1.05,-1.07)$ & $(+4.17,-4.26)$ \\ 
		\end{tabular}
	\end{ruledtabular}
	\caption{\label{fiducialresults}. Median, 68\% and 95\% confidence limit estimates of the three \losen\ parameters from Monte Carlo analysis of the two fiducial models.}
\end{table*}

\begin{figure*}[htp]
	\centering
	\includegraphics[scale=0.38, valign =t]{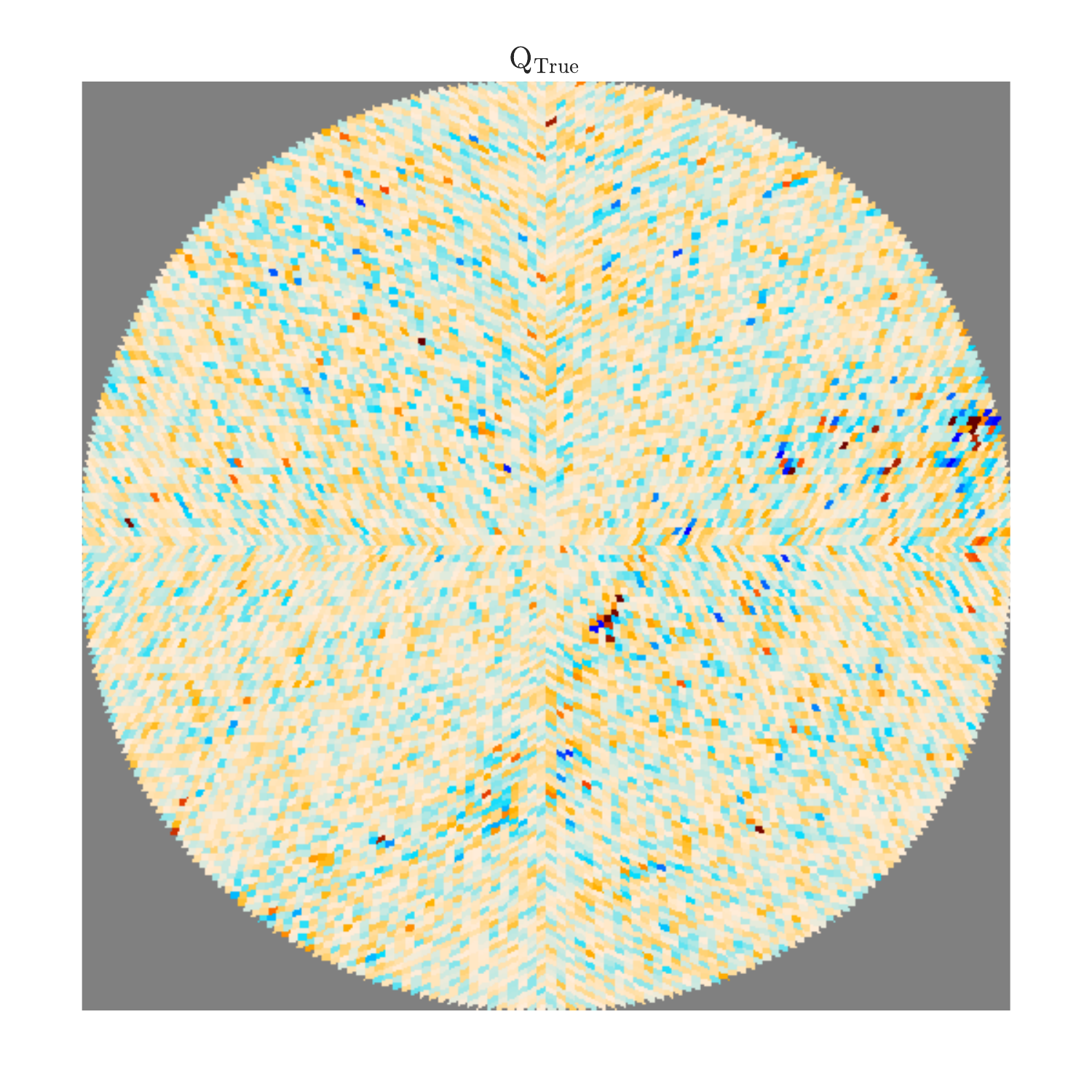}
	\includegraphics[scale=0.38, valign = t]{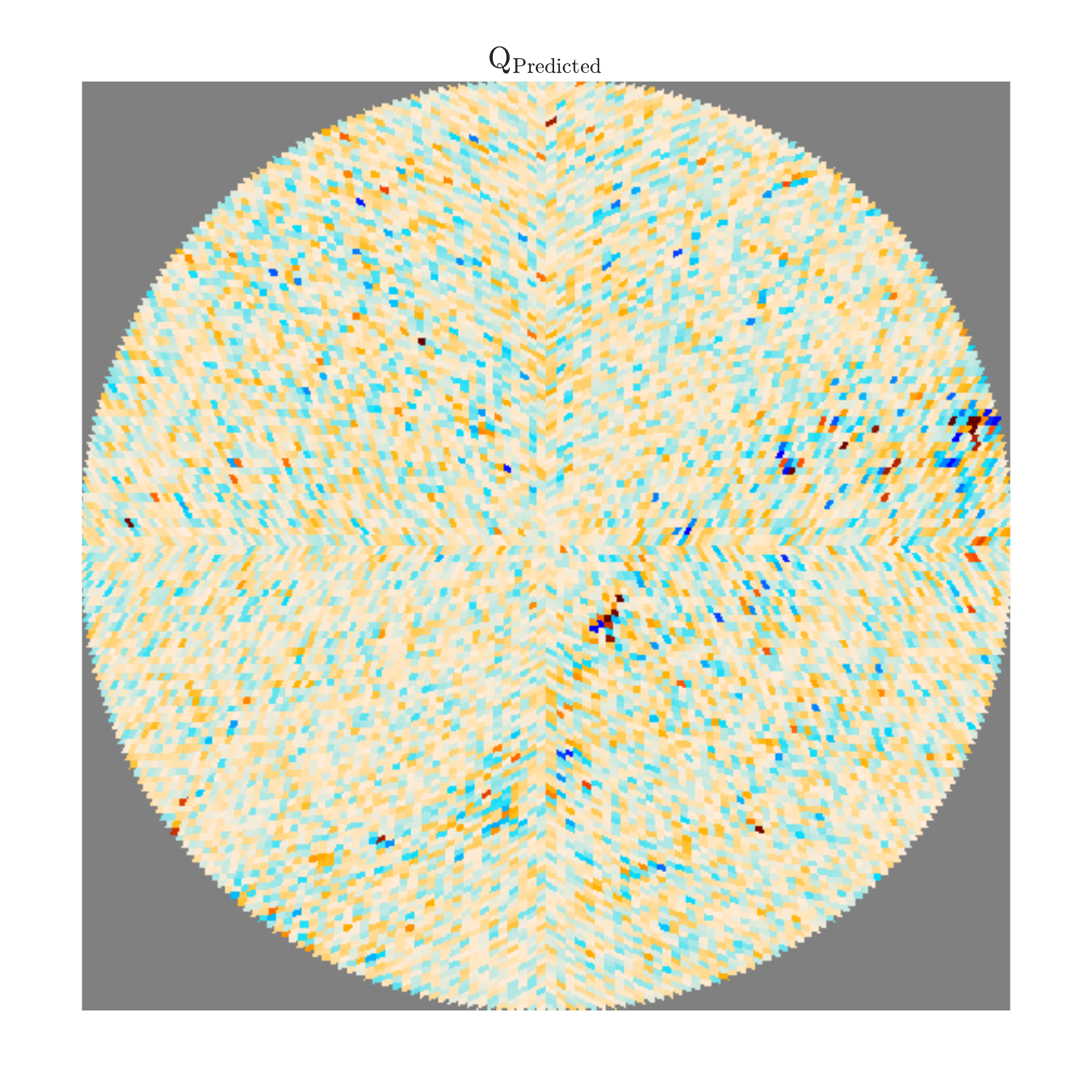}
	\includegraphics[scale=0.38, valign =t]{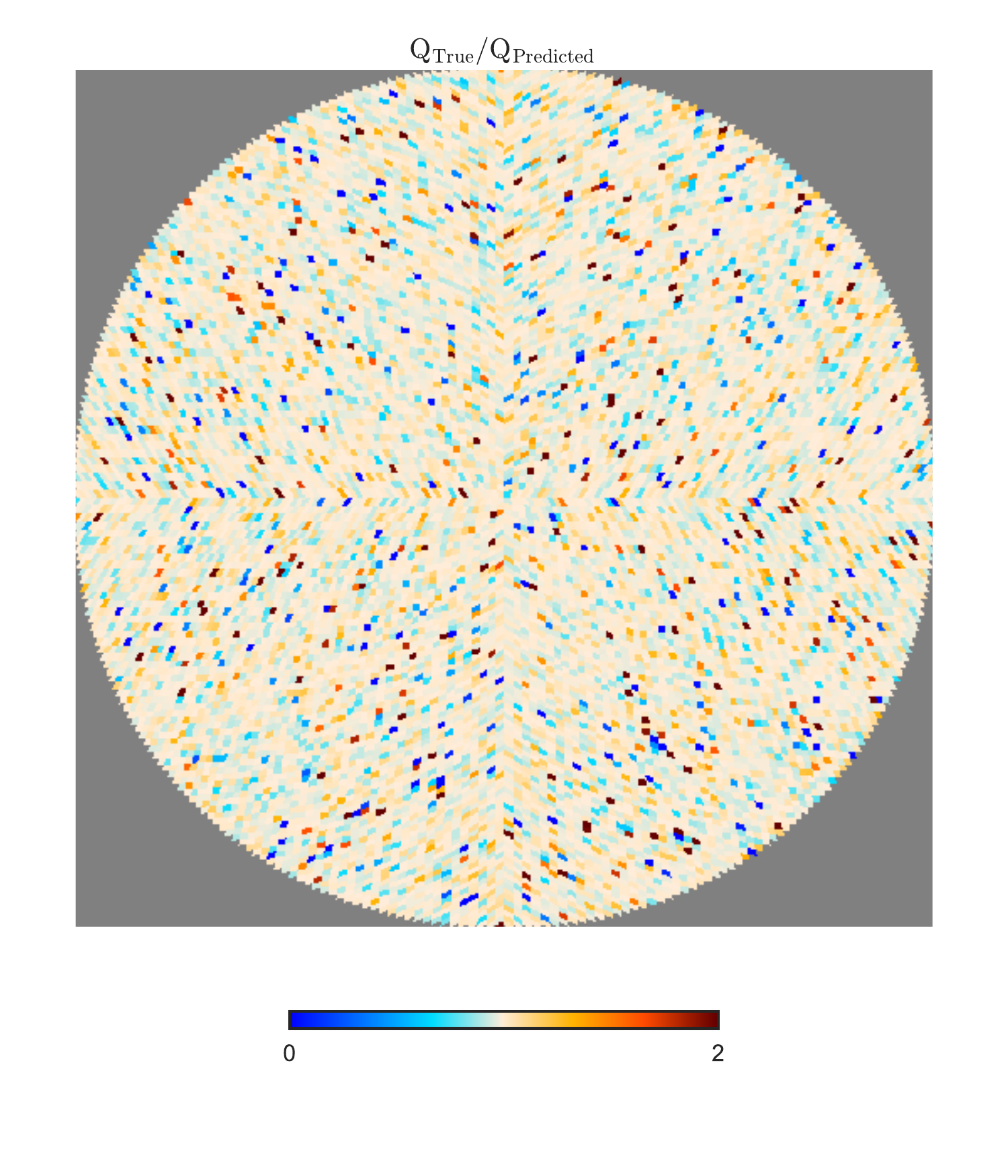}
	\caption{\label{fig:Qratio.pdf} Cartesian projections of stimulated maps of the Stokes Q parameter made using fiducial model 2. Maps are centered on the North Galactic Pole (see \S{\ref{PSdist}} for details). Left: map of true Q Stokes parameters at 150 GHz. Center: map of predicted Stokes Q parameters at 150 GHz, obtained from extrapolating Stokes Q map at 350 GHz down to 150 GHz using \ec{est}. Both maps are unnormalized. Right: map of the ratio of the true Stokes Q parameters to the estimated Stokes Q parameter at 150 GHz. }
\end{figure*}

\begin{figure*}[htp]
	\centering
	\subfigure[Poisson Distribution Model.]{\includegraphics[scale=0.45]{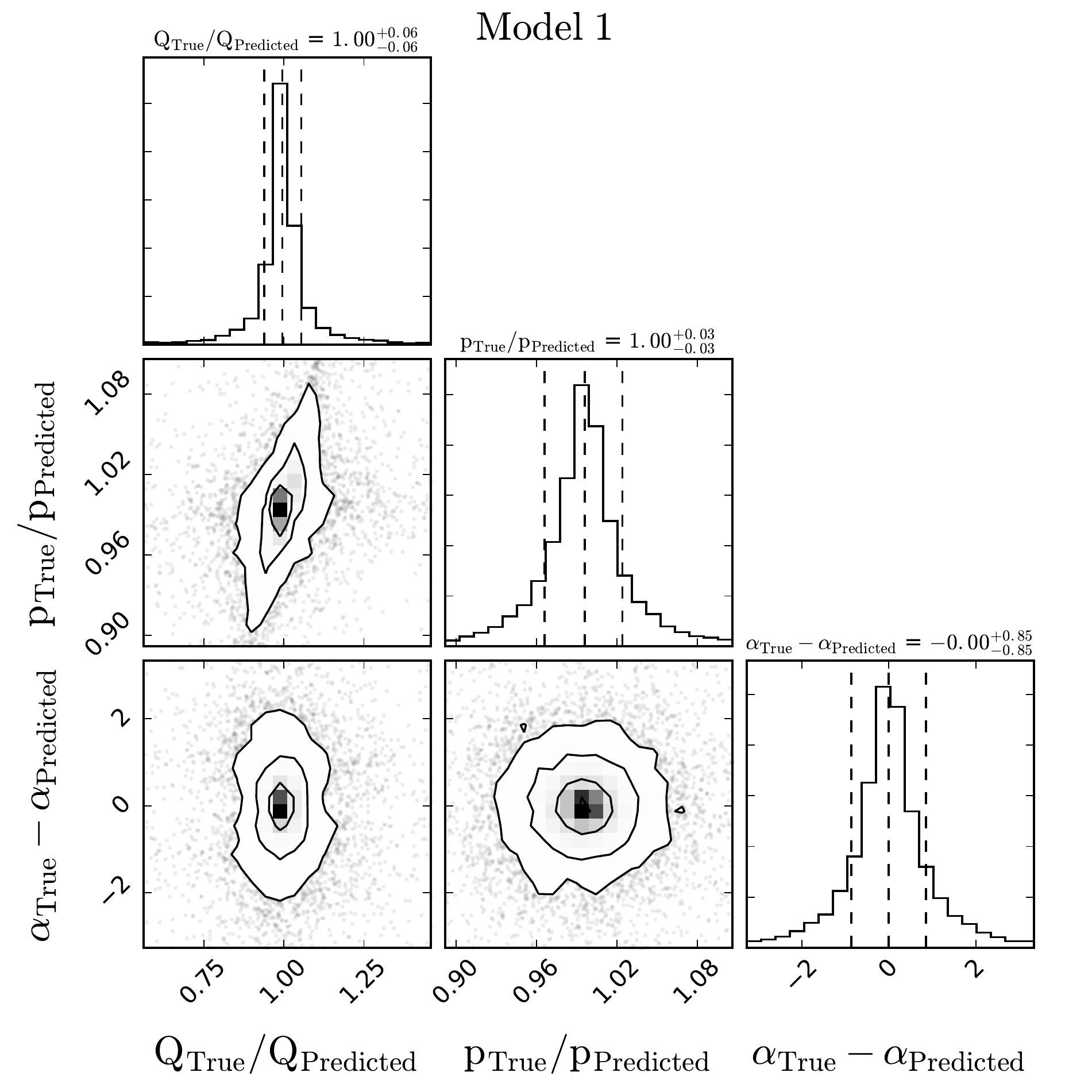}}\quad
	\subfigure[Pan-STARRS 1 Reddening Model. ]{\includegraphics[scale=0.45]{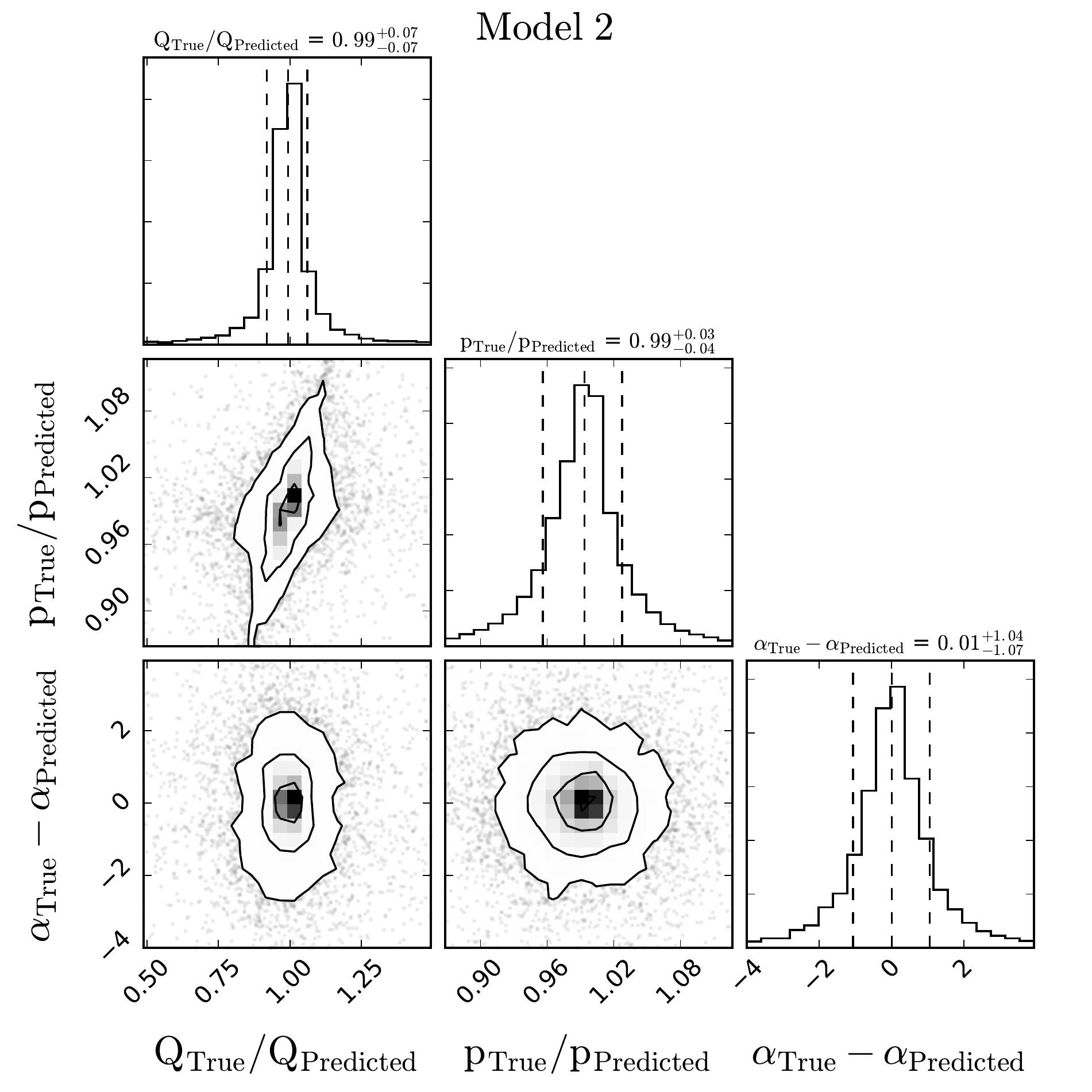}}
	\caption{\label{fig:corner.pdf} Full projected and marginal distributions of \losen\ quantities. In the marginal distribution plots, the median (50th percentile) value and 68th percentile limits are plotted as dashed lines, whose values are stated above each plot. The different 2D projections of the Monte Carlo samples are also directly plotted, with denser regions binned. The contour lines in each 2D projection correspond to the 0.5, 1, 1.5 and 2 $\sigma$ confidence intervals (the 0.5 $\sigma$ line is obscured in some of the plots.)  There appears to be a slight correlation between $p_{\mathrm{True}}/p_{\mathrm{predicted}}$ and $Q_{\mathrm{True}}/Q_{\mathrm{predicted}}$. This is expected, since $p$ has dependencies on the Stokes $Q$ parameter (\ec{pol}). We do not observe a correlation between $\alpha_{\mathrm{True}} - \alpha_{\mathrm{predicted}}$ and any of the other observables, however.}
	\subfigure{\includegraphics[width=2.\columnwidth]{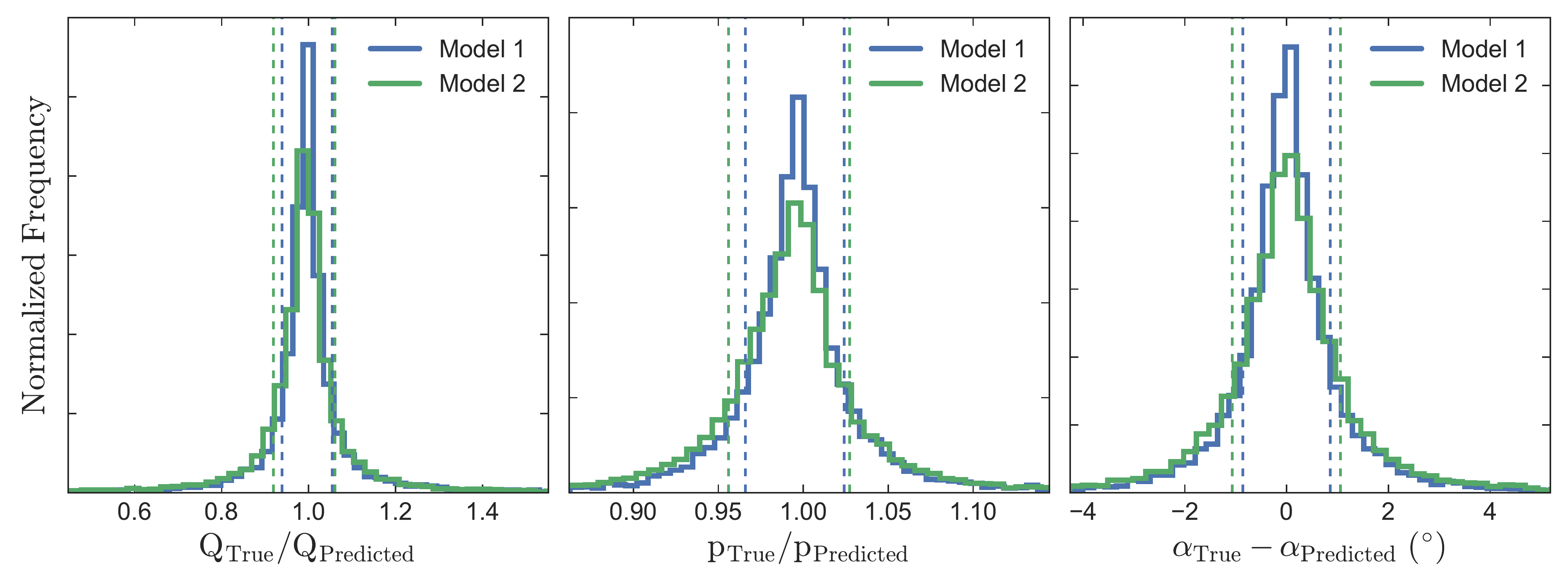}}
	\caption{\label{fig:model_combine.pdf} A comparison of the marginal distributions of \losen\ observables from the two fiducial models, with corresponding input parameters specified in Table \ref{fiducialtable}. As in \Rf{corner.pdf}, the dashed lines corresponds to the 68th percentile confidence intervals.  Both models produce very similar distributions, with model 2 producing samples with slightly wider confidence intervals than model 1. The exact values of the confidence intervals are given in both Table \ref{fiducialresults} and \Rf{corner.pdf}.}
\end{figure*}

We start by using our fiducial models (Table \ref{fiducialtable}) to generate maps of $Q,U$ at 150 and 350 GHz. \Rf{Qratio.pdf} shows examples of the simulated Stokes $Q$ maps made using fiducial model 2. From these maps, we extract:
\begin{itemize}
\item the ratio of true to the predicted Stokes parameter $Q$ at 150 GHz
\item the ratio of true to the predicted polarization fraction $p$ at 150 GHz
\item the difference between the true and predicted 150 GHz polarization angle $\alpha$
\end{itemize}

For our two fiducial models, the results from our Monte Carlo analysis are given in Figs.~\rf{corner.pdf} and \rf{model_combine.pdf} and Table~\ref{fiducialresults}. The distribution profile for both models are very similar, with only a slight difference in the scatter between the two models. We draw the following conclusions from our fiducial models:

\begin{enumerate}
	\item The \losen\ does not bias the estimate of $Q$, $p$ or $\alpha$ in any systematic manner, since the median value for each parameter is consistent with the expected values of those parameters if there were no \losen.
	\item In these fiducial models, the PDF for each parameter is relatively symmetric about the median value, with the upper and lower 68th and 95th percentile limits being comparable in width to each other.
	\item The \losen\ is non-Gaussian, with a cusp at the median value and a longer tail compared to a Gaussian distribution. This can be seen in the difference between the 68th and 95th percentile confidence limits for each of the \losen\ parameters; the width of the 95th percentile confidence limits are 4-5 times the width of the 68th percentile confidence limits, contrary to the expectations for a Gaussian distribution.
	\item Model 2 produces results in a slightly larger \losen\ than model 1, as can be seen in the slightly larger width of its 68th percentile confidence intervals for all 3 parameters (e.g. \Rf{model_combine.pdf}). 
\end{enumerate}

These slight differences notwithstanding, both fiducial models predict 68th percentile statistical uncertainties on the of order 7$\%$ in $Q$ ($U$), 3$\%$ in $p$ and 1$\degree$ in $\alpha$, and 95th percentile uncertainties of order 50$\%$ in $Q$ ($U$), 10$\%$ in $p$ and 4$\degree$ in $\alpha$ per line-of-sight due to \losen. We will compare this error contribution to estimates of the total extrapolation uncertainty reported by the Planck Collaboration in \S{\ref{Planckcomparison}} to ascertain the importance of this effect relative to other sources of error. 

However, it is important to recognize the implication of the longer tail in the distribution on the \losen: The majority of the lines-of-sight have small deviations from the estimator (\ec{est}); however, there is a small population of sightlines where the polarization properties deviate significantly from the estimator when extrapolating between two frequencies, resulting in large mis-estimation of the dust polarization properties at the target frequency of the CMB experiment. It is these particular sightlines that are the greatest cause for concern in polarized dust foreground separation in CMB experiments, since the polarization properties for these lines-of-sight at 350 GHz are not predictive of that at 150 GHz. Masking these particular lines-of-sight will significantly improve constraints on this source of uncertainty. In \S{\ref{future}}, we discuss strategies to account for these non-predictive lines-of-sight.

\subsection{Extension of Fiducial Model: Cloud Number}{\label{fiducialext}}

\Sfig{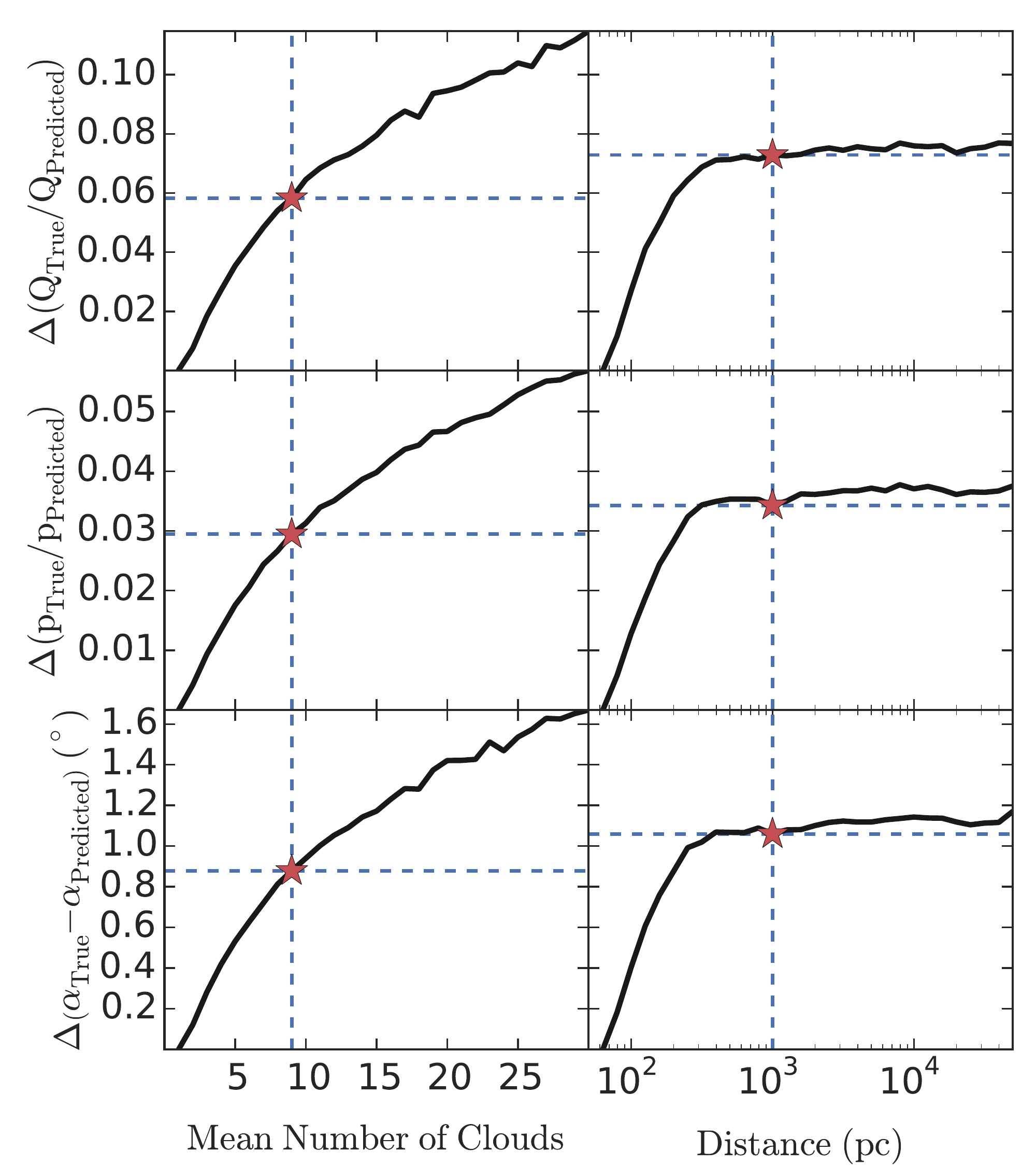}{The \losen\ $\Delta \chi = (\chi_{84\%} - \chi_{16\%}) / 2 $, for the 3 \losen\ parameters. The left column corresponds to model 1 for various values of the mean number of cloud, while the right column corresponds to model 2 at various cumulative distance bins. Stars indicate values for the fiducial models we describe in \S{\ref{resfid}}.}

The largest unknown quantity in our modeling is estimating how many contributing clouds there are along a line-of-sight. Here, we extend our fiducial models to characterize how the \losen\ scales with the number of contributions along a line-of-sight. For model 1, we vary the mean number of clouds per kpc from 1 to 30 in the Poisson distribution of number of clouds along a line-of-line. For model 2, we extend the cumulative number of reddening distance bins we include in our Monte Carlo analysis out to the furthest distance bin corresponding to a distance of $\sim63$ kpc. 

For each variation in the number of contributions, we keep all other parameters fixed as given in Table~\ref{fiducialtable}, except for the distributions of dust temperature $T$ and polarization fraction $p$. For the $T$ and $p$ distributions, we refit the 3D temperature and polarization fraction PDF in order to reproduce the observed line-of-sight distributions for each variation in the number of contribution along a light-of-sight. The fitted 3D temperature and polarization fraction distribution parameters for each cloud distribution model are given in Table \ref{TPtable}. Finally, we parametrize the extrapolation uncertainty as half of the width spanned by the 68th percentile confidence limits, $\Delta \chi = (\chi_{84\%} - \chi_{16\%}) / 2 $, where $\chi \in \{Q_{\mathrm{True}}/Q_{\mathrm{predicted}},\, p_{\mathrm{True}}/p_{\mathrm{predicted}},\, \alpha_{\mathrm{True}} - \alpha_{\mathrm{predicted}} \} $. 

\Rf{combine.pdf} shows the results of this analysis. The left plots show the \losen\ using model 1 for various values of the mean number of clouds along a line-of-sight, ranging from 1-30. In this model, the \losen\ increases monotonically with the mean number of clouds. However, the rate of increase in \losen\ appears to fall off with a larger number of clouds. For model 2, the \losen\ flatten off much more significantly after $\sim$1 kpc. The leveling off is likely due to the reddening in majority of the sight lines in the $30\degree$ radius region centered on the North Galactic Pole falling off after that distance bin. 

We conclude from this analysis that the two fiducial models are relatively consistent with each other, i.e. the Pan-STARRS 1 reddening map is consistent with a Poisson distribution model with a average of about $10$ clouds along each line-of-sight. This consistency check supports our choice of fiducial values for the cloud number distribution.  

\subsection{Systematics}

In \S\ref{resfid}, we explored the line-of-sight extrapolation noise levels from our fiducial models, and in \S\ref{fiducialext}, extensions of our fiducial model for different distributions of number of contributing clouds along a line-of-sight. Here, we explore the various possible systematic uncertainties that may potentially bias our result. We investigate the dependence of the \losen\ on the distribution of input parameters as well as possible biases that may result from a specific choice of the angular resolution scale. 

\subsubsection{\label{senseinput} Input Parameter Distributions}
\Sfigbig{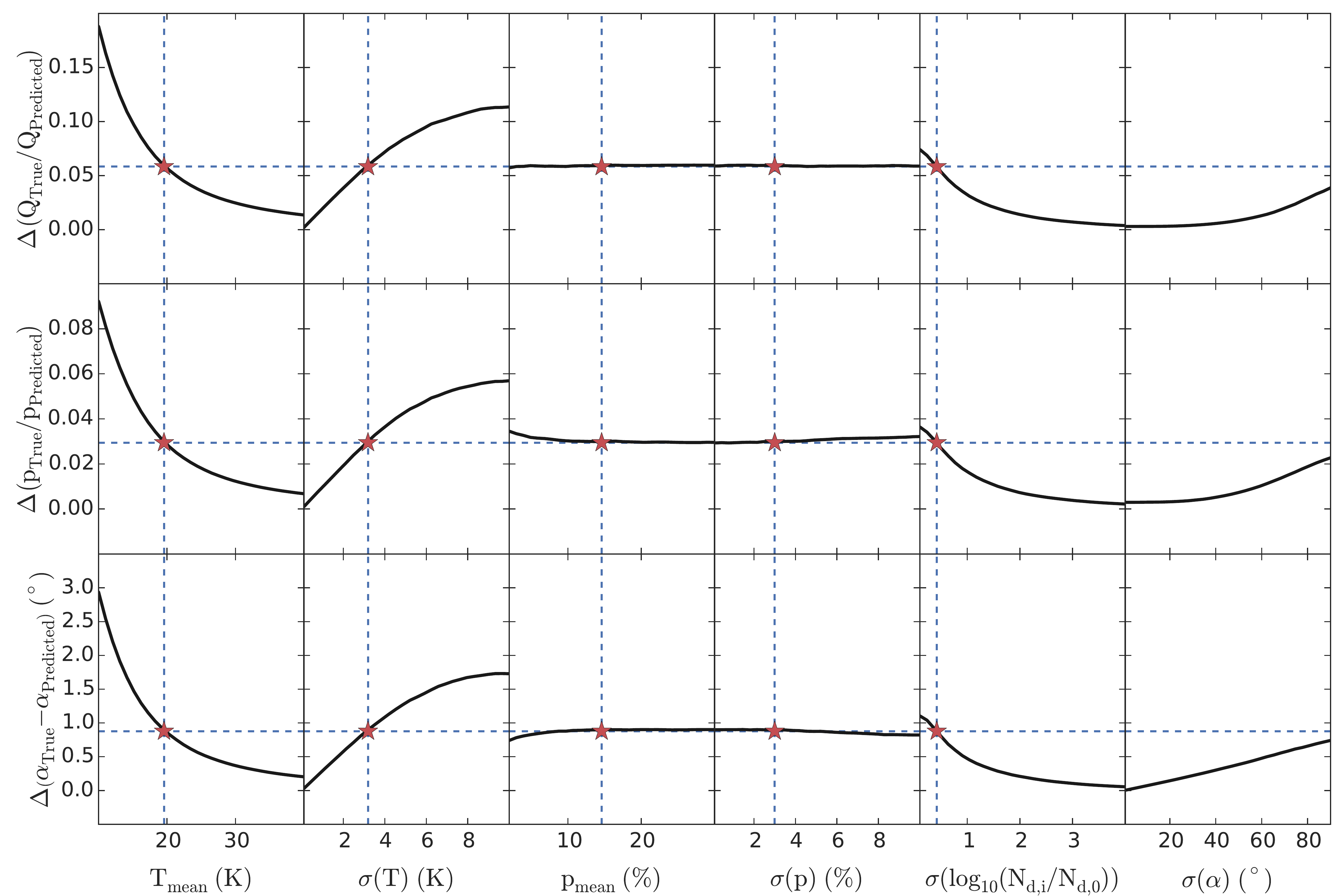}{Extrapolation noise as a function of various input distributions using model 1, with stars indicating fiducial values. The line-of-sight extrapolation noise has no dependence on the mean values of $\log_{10}(N_{d,i}/N_{d,0})$ or $\alpha$ in the distribution, and so those dependencies are not plotted. In the rightmost column, we investigate the possible effect of correlations in the polarization angles of clouds along a line-of-sight on the extrapolation noise by changing the polarization angle distribution from the fiducial choice of a uniform random distribution to a Gaussian distribution around an arbitrary mean angle, and plot the extrapolation noise as a function of the standard deviation of the polarization angle distribution.} 

We investigate the dependence of the line-of-sight extrapolation noise on the input parameter distributions by varying the value of a single input parameter while fixing the remaining parameters and characterizing how the extrapolation noise varies with the parameter. In this analysis, we use model 1 and vary individual distribution parameters for $T$, $p$, $N_{d}$ and $\alpha$ while fixing the remaining parameters at the fiducial values (Table \ref{fiducialtable}). We use model 1 in this analysis for expediency, as the parametrization of the cloud number and column density in this model is easy to modify. However, we expect the effect of variations in the input distributions on the line-of-sight extrapolation noise to hold for any general model of dust distribution, since the trends reflect the underlying physical mechanisms governing the extrapolation noise and not the distribution of dust clouds. 

The results are shown in \Rf{error_combine.pdf}, with fiducial values highlighted for comparison. The first two columns plot variations in the line-of-sight extrapolation noise with changes in the cloud temperature distributions. The \losen\ has a strong dependence on the form of the cloud temperature distribution, as there are obvious variations in the extrapolation uncertainty when the mean and standard deviation of the temperature distribution are modified. A higher mean cloud temperature leads to lower extrapolation noise, e.g. from $\Delta(Q_{\mathrm{True}}/Q_{\mathrm{predicted}})=0.3$ to 0.02 when the mean temperature increases from 10 K to 40 K. The physical reason for this is that the blackbody spectral radiance is closer to the Rayleigh-Jeans regime at higher temperatures in these frequencies, and therefore, the true frequency-scaling relation (e.g. \ec{losratio}) increasingly converges to the estimator \ec{est}. On the other hand, an increase in the standard deviation of the cloud temperature distribution leads to an increase in the line-of-sight extrapolation noise. This is also consistent with our physical understanding of the extrapolation noise, since a larger variation in temperature between clouds leads to a greater mismatch in the frequency-scaling between dust clouds along a line-of-sight. 

The third and fourth columns show how the line-of-sight extrapolation noise varies with the distribution of polarization fractions. In contrast to the cloud temperature distribution, the \losen depends only weakly on the polarization fraction distribution, as variations in both the mean and the scatter in the polarization fraction distribution do not result in any significant change in the extrapolation noise from its fiducial values. Therefore, we believe that our use of a simplified distribution profile for the cloud polarization fraction is justified. However, this result assumes that the distribution of polarization fractions of dust clouds is independent of frequency, and may not necessarily hold in the case where the polarization fractions of dust clouds vary with frequency.

The fifth column plots the effect on the \losen\ of variations in column densities between dust clouds, parametrized by the standard deviation in the log-normal distribution in the dimensionless ratio $\log_{10}(N_{d,i}/N_{d,0})$, where $N_{d,0}$ is an arbitrary reference column density used for normalization. As a consistency check, we verified that the choice of $N_{d,0}$ is indeed arbitrary by the observation that the extrapolation noise is completely independent of the mean value of $\log_{10}(N_{d,i}/N_{d,0})$, which is determined by the choice of $N_{d,0}$. The plots in column 5 show that larger variations in the cloud column density (parametrized by an increase in the standard deviation in $\log_{10}(N_{d,i}/N_{d,0})$) result in a decrease in extrapolation noise. A physical explanation for this trend is that in a population of clouds with large variations in column densities between clouds, the overall polarized dust SED is dominated by the clouds with the largest column densities, effectively masking out contributions from other clouds along the line-of-sight. This effectively lowers the extrapolation noise, since the extent of \losen\ depends on the contributions of multiple clouds. 

Finally, to investigate the effect on \losen\ of correlations in polarization angles of clouds along a line-of-sight, we change the polarization angle distribution from the fiducial choice of a uniform random distribution to a Gaussian distribution around an arbitrary mean angle, and vary the standard deviation of the distribution. We verified that the choice of mean angle is arbitrary by varying the choice of mean angle and checking that it has no effect on the \losen. We then increase the standard deviation of the polarization angle distribution from $0\degree$ to $90\degree$. As the standard deviation increases, the \losen\ asymptotically approaches that of the fiducial model. The relevant plots are shown in the last column of \Rf{error_combine.pdf}. Large-scale correlations in polarization angles along a line-of-sight effectively decrease the \losen, so our fiducial assumption of a uniform random distribution \textit{overestimates} the \losen.  

In summary, this analysis suggests the following about the input distributions and their effect on our fiducial analysis:

\begin{itemize}
	\item The \losen\ is most sensitive to the temperature distribution of the dust clouds.
	\item The fiducial results are relatively insensitive to the distribution of polarization fractions. Therefore, our analysis is relatively robust with respect to our assumptions about and simplification of the polarization fraction distribution.
	\item Variations in the dust column density has an significant effect on the \losen\ analysis, but the effects are less pronounced compared to the temperature distribution and increases in these variations serve to decrease the \losen.
	\item Our model overestimates the \losen\ if there are large-scale correlations in the polarization angles of dust clouds along the line-of-sight.
\end{itemize}

\subsubsection{Variations in 3D Temperature Distribution} \label{Tsystematics}

\begin{figure*}[htp]
\centering
\subfigure{\includegraphics[width=2.\columnwidth]{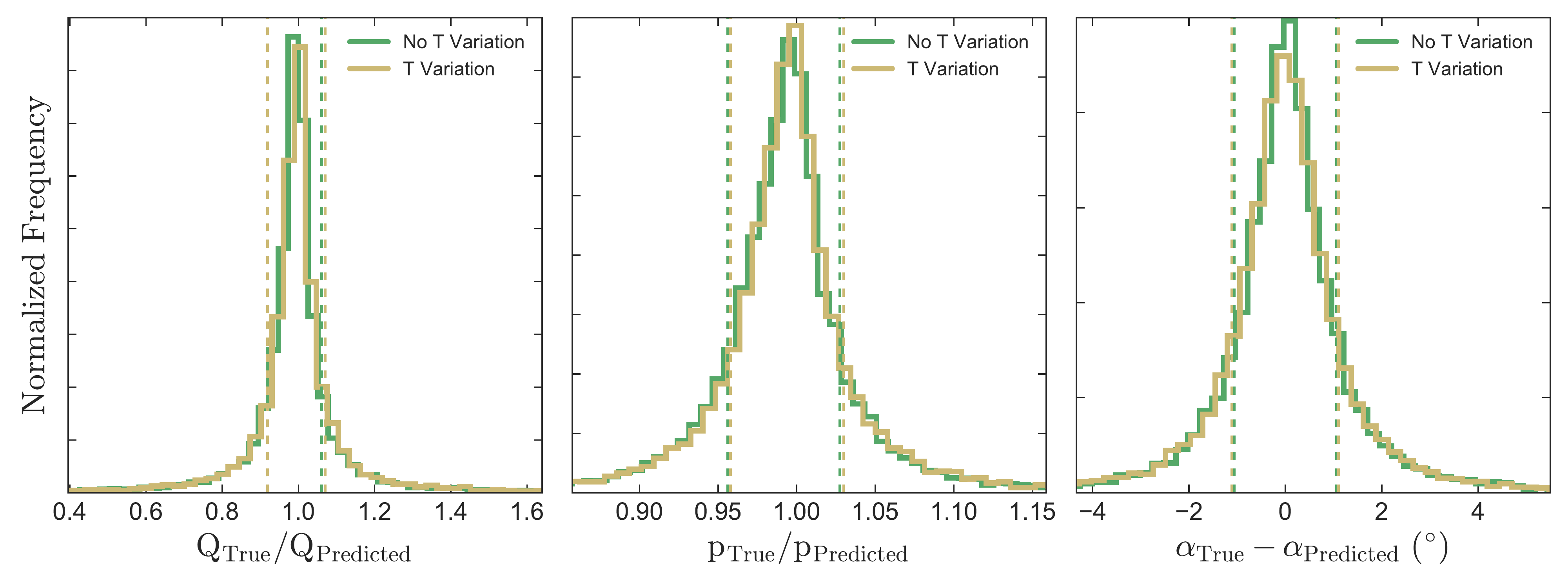}}
\caption{\label{fig:model_T.pdf} A comparison of the marginal distributions of \losen\ observables from (1) the fiducial model 2 (green), where every cloud draws a temperature from the same temperature distribution, and (2) a model in which the mean temperature of the clouds decrease with distance from the Galactic disk (yellow). Dashed lines indicate 68\% confidence intervals for each model. The model used to produce the latter is described in \S{\ref{Tsystematics}}.}
\end{figure*}

Given that the \losen\ is most sensitive to the temperature distribution of the dust clouds, it is worth considering ways in which the 3D dust temperature distribution can be further refined in order to improve the fidelity of our model. One way in which our 3D distribution model can be improved is to account for variations in the temperature distributions for dust clouds at different distances from the Galactic disk. The physical reason for this is that the radiation field from the Galactic disk is the dominant heating mechanism for Galactic dust, and so we expect dust clouds further away from the disk to be systematically cooler than nearby clouds.

Here, we investigate to first-order the effects of a systematic variation in dust temperature with distance by considering a model where, instead of drawing a temperature for each dust cloud from the same universal temperature distribution, dust clouds at different distances draw temperatures from different temperature distributions, where the mean temperature of each distribution decreases as a function of distance. Model 2 is a natural fit for this study, because each reddening contribution is associated with a distance bin. 

For simplicity, we consider a toy model where we use the fiducial 3D Gaussian temperature distribution (Table \ref{fiducialtable}), scaling only the mean temperature of the distribution such that it decreases with each distance bin. We follow the toy model described by Tassis, Pavlidou and Kylafis \cite{Tassis2016a}, where we assume that spherical dust clouds are situated at different distances $h$ above the center of the Galactic plane, and are heated only by a uniform disk of stars within the plane. Assuming each cloud is at thermal equilibrium, absorbing the same fraction of the incident flux from the stellar disk and emitting thermally, the temperature will decline with $h$ as 
\ba \eql{Tvareqn}
T \propto \ln\left(1+\left(\frac{R_{\mathrm{disk}}}{h}\right)^2\right)^{1/4}
\ea
where $R_{\mathrm{disk}}$ is the radius of the stellar disk. Here, we set $R_{\mathrm{disk}}$ to 13.5 kpc, following \cite{Kennicutt2012}. We then fit for the temperature of the nearest distance bin in Model 2 so as to reproduce the observed line-of-sight integrated temperature distribution of $T=19.7\pm1.4$ K. The best fit model has a mean temperature of $\sim$20.5K for the nearest distance bin of 0-63 pc, which falls off to $\sim$17.2K at 1 kpc. We then calculated the \losen\ parameters for this model and compared it to the fiducial model, where the temperature distribution of the dust clouds does not depend on distance. 

The results, plotted in \Rf{model_T.pdf}, show that this model, where the mean cloud temperatures decrease non-linearly with distance from the Galactic plane, does not produce a significant difference in the \losen, compared to the fiducial model. However, this result may not necessarily hold for more sophisticated and physically representative models of the distribution of dust temperatures, and we caution against broadly generalizing from this result.

\subsubsection{Angular Resolution of Pan-STARRS 1 Dust Map}

\Sfigbig{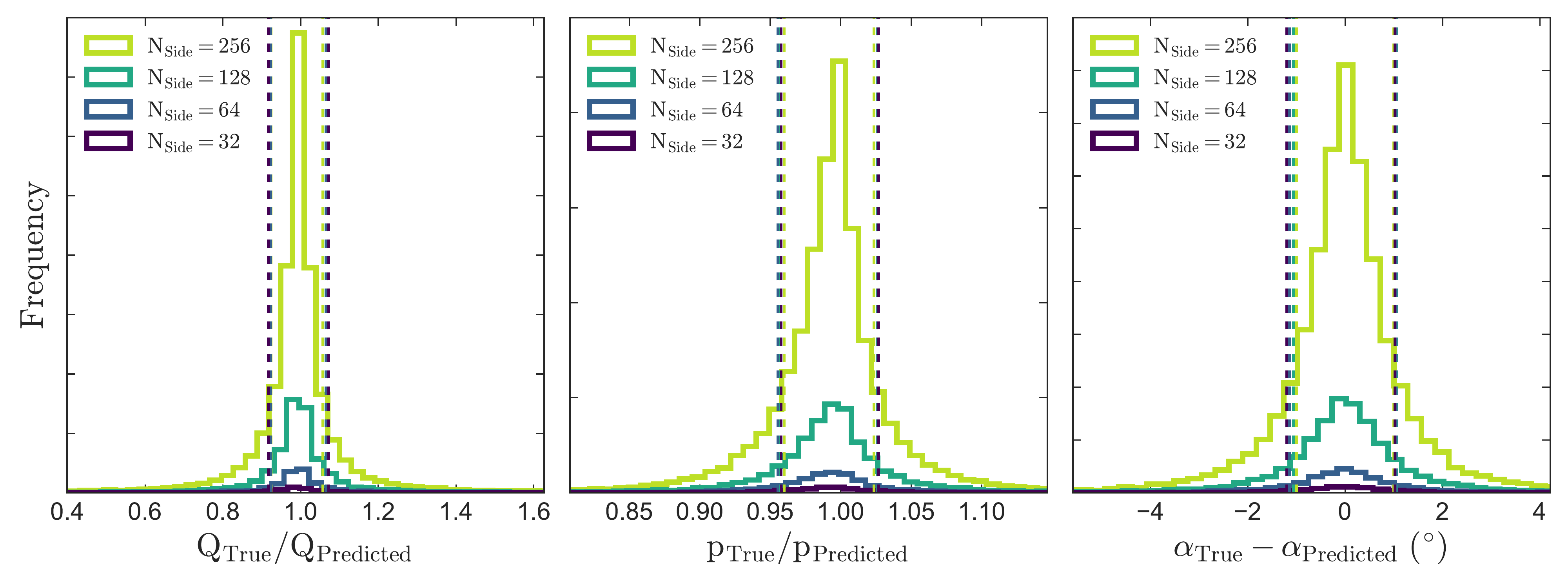}{Unnormalized marginal distribution of \losen\ parameters at various $N_{\mathrm{side}}$ angular resolution using model 2 and fiducial distributions (Table \ref{fiducialtable}). Dashed lines of the same color indicate 68$\%$ confidence intervals for each of the four distributions. For each $N_{\mathrm{side}}$ value, the dust reddening map was downsampled from a native resolution near the Galactic Pole of $N_{\mathrm{side}}=256$ down to its target resolution by averaging the reddening of all the native resolution pixels in each target resolution pixel. Other input parameters were drawn from the same fiducial distribution.}

Here, we explore potential systematic uncertainties associated with the native and downscaled angular resolution of the Pan-STARRS 1 dust maps used to infer dust column densities. The native resolution of the Pan-STARRS 1 map in the $30\degree$ radius region centered on the north Galactic pole is 13.7', which we degrade to 27.5' (corresponding to $N_{\mathrm{side}} = 128$) by assigning to each downsampled pixel the mean reddening value of the children pixels within it. A concern is that the loss of granularity in variations in the dust cloud reddening on small angular scales could bias our analysis. 

To investigate this potential systematic uncertainty, We calculate the \losen\ at resolutions of 13.7', 27.5', 55' and 110' (corresponding to HEALPix $N_{\mathrm{side}}$ resolutions of 256, 128, 64, and 32 respectively), degrading the map by assigning to each pixel at the target resolution the mean reddening value of the children pixels at $N_{\mathrm{side}}=256$ within it. The temperature, polarization fraction and polarization angle were sampled independently from the same fiducial distribution for each target pixel. 

\Rf{modelnside.pdf} shows the \losen\ when the variation in dust cloud reddening is effectively smoothed over different pixel resolutions. The plots show an increase in the $68\%$ confidence interval with increasing  $N_{\mathrm{side}}$ resolution, but the dependence is very weak, implying that \losen\ is not particularly sensitive to the granularity of the dust clouds itself.

\subsubsection{Map Smoothing and Pixel Angular Scale Degradation} \label{mapsystematics}

\Sfigbig{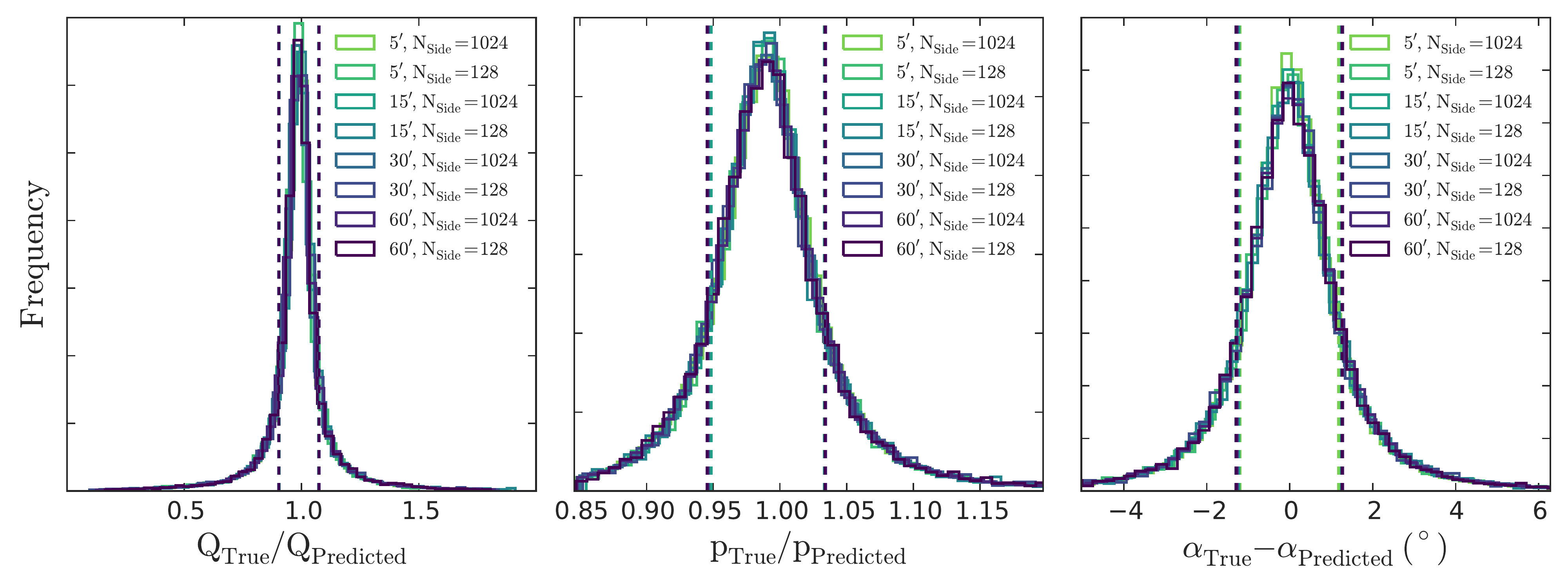}{Normalized histograms of the three Extrapolation error parameters for various degrees of smoothing at angular resolutions of 3.4' ($N_{\mathrm{side}} =1024$) and when degraded to a pixel angular resolution of 13.7' ($N_{\mathrm{side}} =128$), the angular resolution of the pixel used in our fiducial analysis as well as the Planck analysis \cite{Ade2015b}.  Vertical dashed lines indicate 68$\%$ confidence intervals of histograms of the same color.}

A potential concern with our analysis is that the \losen\ may depend on the angular resolution of our mock maps, and that systematic biases can arise from the choice of angular resolution scale used in our analysis. Characterizing these systematics are important for a accurate comparison of our fiducial results with the observed extrapolation uncertainties reported by the Planck collaboration, because in the Planck analyses, polarization maps were smoothed from their native angular resolution (e.g. 5' for 353 GHz) to $1\degree$ FWHM resolution, and the pixel resolution of those maps were degraded to e.g. $N_{\mathrm{side}} = 256$ in \cite{Ade2015} and $N_{\mathrm{side}} = 128$ in \cite{Ade2015b}. In this section, we investigate the effect of (1) smoothing the polarization maps and (2) downsampling the HEALPix pixel resolution of the polarization maps on the \losen.

We investigate the effects of smoothing by generating Stokes $I$, $Q$ and $U$ maps at 150 GHz and 350 GHz at an angular resolution of 3.4' (corresponding to $N_{\mathrm{side}} = 1024$) using fiducial model 2.  The maps at those two frequencies are then smoothed with a Gaussian beam of FWHM 5', 15', 30' and 60'. The \losen\ was then calculated for each pixel of the smoothed maps, using estimator \ec{est} with an inferred temperature obtained by fixing $\beta=1.59$ and fitting temperature parameter in the estimator to the ratio of the line-of-sight specific intensities $I_\nu$ at 150 GHz and 350 GHz for each pixel. Each of these smoothed maps are then degraded to a pixel resolution of 13.7' (corresponding to $N_{\mathrm{side}} = 128$) by assigning each $N_{\mathrm{side}} = 128$ pixel the mean Stokes $I$, $Q$, and $U$ parameters of the $N_{\mathrm{side}} = 1024$ pixels within it. 

\Rf{modelsmoothds.pdf} show superimposed histograms of \losen\ parameters of the eight maps with various smoothing and pixel angular scale degradation. The vertical lines indicate $68\%$ confidence intervals. We find that the $68\%$ confidence intervals of the \losen\ does not vary significantly, either with the degree of smoothing or the degradation of the pixel angular resolution scale. Therefore, map smoothing or pixel resolution scale degradation does not appear to bias or change the systematic uncertainty caused by the \losen.

\section{Comparison with Planck Uncertainty}\label{Planckcomparison}

We want to determine how significant \losen\ is compared to the total extrapolation uncertainty reported by the Planck experiment from cross-correlation analyses of the dust polarization data at intermediate latitude sky patches \cite{Ade2015b}. The Planck experiment currently has the most sensitive all-sky maps in these frequencies, and these maps have been used in the most recent B-modes analyses by the joint BICEP2/Planck collaboration \cite{Ade2015c,Ade2016}, which makes them particularly relevant to this study. In their analysis, the Planck collaboration determined the spectral indices of polarization, $\beta$, for 400 sky patches of $10\degree$ radius at intermediate latitudes at a HEALpix resolution of $N_{\mathrm{side}} = 128$, and reported a mean spectral index with $1\sigma$ dispersion of $\beta = 1.59\pm 0.17$. This observed $1\sigma$ dispersion in $\beta$ represents the {\it total} observed extrapolation uncertainty due to extrapolation, and includes both the Planck HFI instrument noise  and \losen.

Upcoming CMB experiments will be able to reduce instrumental noise by virtue of better sensitivities and angular resolution. However, the \losen\ represents a component of the intrinsic astrophysical foreground uncertainty in the polarized dust foreground separation technique that may have to be accounted for by future CMB experiments. Here, we estimate the contributions of the \losen\ as well as the Planck instrument noise and compare it to the total extrapolation uncertainty. 

\subsection{Line-of-Sight Dust Extrapolation Noise}\label{contributionDEN}

We consider the impact of \losen\ by performing the following analysis, using mock Stokes $Q$ and $U$ maps of only the polarized dust emission, generated using our Pan-STARRS 1 fiducial model (model 2) at various frequencies for a 30\degree\ radius region centered on the Galactic Pole. We emulate the Planck analysis \cite{Ade2015b} by generating 400 mock sky patches, each comprising 1000 independent pixels. The Stokes $Q$ and $U$ parameters for each pixel were randomly sampled without replacement from the mock Stokes maps. For each sky patch, we then calculate the polarization cross-correlation coefficient at frequency $\nu$, $\alpha_\nu$, by minimizing the $\chi^2$ expression using the 353 GHz map as a template\footnote{This expression differs slightly from the Planck analysis (Eq. (13) of \cite{Ade2015b}) in that we omit fitting for a constant local mean offset between the different frequency maps, as our mock maps do not contain that systematic effect.}:
\ba  \eql{cceqn}
\chi^2 = \sum_{i=1}^{1000} \left[Q_i(\nu)-\alpha_{\nu} Q_i(353\ \mathrm{GHz})\right]^2 \\ + \left[U_i(\nu)-\alpha_{\nu} U_i(353\ \mathrm{GHz})\right]^2
\ea
where the sum is over every pixel in the sky patch. The cross-correlation coefficients are then fitted with the usual modified blackbody parametrization\footnote{The fiducial Planck analysis was done in units of thermodynamic temperature ($\unit{K_{CMB}}$), and so the parametrization they used (Eq. (19) of \cite{Ade2015b})  has to account for instrumental color correction and unit conversion factors. Here, we are in units of $\unit{MJy\,Sr^{-1}}$, so we omit these factors}, 
\ba \eql{MBBfit}
\alpha_{\nu} \propto B(\nu, T)\,\nu^{\beta} 
\ea
We can then deduce the spectral index $\beta$ for the sky patch from the cross-correlation coefficient, $\alpha_{\nu}$, given an independent estimate of the dust temperature of the sky patch. Each sky patch produces an estimate of $\beta$, and the $1\sigma$ dispersion in $\beta$ across the 400 mock sky patches provides an estimate of the error due to \losen, which we compare with the total extrapolation uncertainty of $\Delta \beta = 0.17$ reported by the Planck experiment.

To estimate the impact of \losen\ on $\beta$, we implement two different methods to estimate the dispersion in $\beta$ when extrapolating the dust polarized SED from 353 GHz to 150 GHz. The first, more straightforward method directly calculates the cross-correlation coefficients at 150 GHz, $\alpha_{150}$, from mock Stokes $Q$ and $U$ maps at 150 GHz and 353 GHz for the 400 sky patches. We then fit the modified blackbody spectrum \ec{MBBfit} to each cross-correlation coefficient using the mean dust temperature of 19.6 K reported by Planck to deduce the spectral index for each sky patch. Over the 400 mock sky patches, we obtained a 68th percentile dispersion in $\beta$ of $\Delta \beta = 0.006\pm 0.0003$, where the error is obtained from bootstrapping. This error is $\sim4 \%$ of the total extrapolation uncertainty reported by Planck.

Our second method more closely emulates the fiducial Planck analysis by inferring $\beta$ not directly from $\alpha_{\nu}$, but from the color ratio R(100,217,353)\footnote{This parametrization is used in the Planck analysis because the difference in the cross-correlation coefficients (in units of $\unit{\mu K_{CMB}}$) subtracts the achromatic CMB contribution, while the fraction removes normalization terms. Our mock maps do not contain these contributions; however, \losen\ varies with frequency, so this parametrization will produce a different estimate of the \losen.}, where $R$ is a combination of cross-correlation coefficients
\ba \eql{colorratio}
R(\nu_0,\nu_1,\nu_2) = \frac{\alpha_{\nu_2} - \alpha_{\nu_0}} {\alpha_{\nu_1} - \alpha_{\nu_0}} 
\ea 
 Following the fiducial Planck analysis, we generate mock maps at 100, 217 and 353 GHz, and calculate the color ratio $R$ for each pixel. We then infer $\beta$ by fitting $R$ for each sky patch, using the same mean dust temperature of 19.6 K. Using this method, we obtained a slightly higher 68th percentile dispersion in $\beta$ of $\Delta \beta = 0.007\pm 0.0003$. 

\subsection{Planck HFI Instrument Noise}\label{contributionPDS}

To obtain an estimate of the contribution from the Planck HFI instrument noise to the total extrapolation uncertainty, we conduct the following rudimentary analysis: We first generate signal-only maps at multiple frequencies, such that the dust polarization SED scales with frequency exactly as a modified blackbody (\ec{est}) with a uniform temperature and frequency. We then generate instrument noise maps at those frequencies, and add the noise component to the signal-only maps to create a ``signal + noise" polarized dust emission map where the only uncertainty in the polarized dust SED comes from the instrument noise. We then emulate the Planck analysis \cite{Ade2015b} to infer the dust spectral index $\beta$ for $\sim400$ sky patches of 10\degree\ radius. Any scatter in the inferred dust spectral index from these sky patches would arise entirely due to the instrument noise. Hence, we consider the dispersion in $\beta$ from this analysis an approximate estimate of the contribution from Planck instrument noise to the total extrapolation uncertainty. 

We use the polarized dust emission map at 353 GHz from the Planck 2015 astrophysical component analysis\footnote{Available publicly as part of the Planck Public Data Release 2: http://irsa.ipac.caltech.edu/data/Planck/release$\_$2/all-sky-maps/foregrounds.html} as a proxy for the signal-only component of the thermal dust polarization map at 353 GHz. Following the Planck analysis \cite{Ade2015b}, we first smooth the map to a resolution of 1\degree\ and downsample the map to a HEALPix resolution of $N_{\mathrm{side}} = 128$. We then use this signal template to generate signal maps of thermal dust polarization at lower frequencies (e.g. 217 GHz) by first converting the maps from units of antenna temperature, $\unit{K_{RJ}}$, to units of $\unit{MJy\,Sr^{-1}}$ and then scaling the signal of each pixel with frequency using the estimator \ec{est}, assuming a uniform temperature of $T=19.6\,$K and  polarization spectral index of $\beta = 1.59$. 

To generate the Planck instrument noise maps, we used the difference in the Planck half-mission frequency maps as a proxy for the instrument noise at different Planck HFI frequency bands. We then combine the signal and noise maps by summing the signal and the noise components from the two maps for each pixel, converting both maps to units of $\unit{MJy\,Sr^{-1}}$ beforehand for unit consistency. 

We then infer $\beta$ from these noisy maps in a similar fashion as the Planck analysis \cite{Ade2015b}. First, we divide the sky map into patches of 10\degree\ radius centered on HEALPix pixels at a resolution of $N_{\mathrm{side}} = 8$. Emulating the Planck analysis, we consider 488 sky patches centered at intermediate Galactic latitudes of $10\degree< |b|<60\degree$. For each sky patch, we obtain the cross-correlation coefficient at various frequencies, following the same $\chi^2$ minimization procedure as \S{\ref{contributionDEN}} (\ec{cceqn}). 

The cross-correlation coefficients can then be used to infer the spectral index for each sky patch. Using the cross-correlation coefficient at 217 GHz, $\alpha_{217}$, to directly infer the spectral index, we obtain a $1\sigma$ dispersion in $\beta$ of $0.19 \pm 0.03$ from these maps, where the error is from bootstrapping. We also calculated the color ratio R(100,217,353) (\ec{colorratio}) for each sky patch and inferred $\beta$ from those values, obtaining a $1\sigma$ dispersion in $\beta$ of $0.22 \pm 0.03$ from these maps. These estimates of the contribution from Planck instrument noise to the scatter in $\beta$ are consistent with the total observed extrapolation uncertainty of $\Delta \beta = 0.17$, suggesting that the Planck HFI instrument noise can account for most of the total extrapolation error reported by the Planck experiment. 
 
\subsection{Intrinsic Variation in Spectral Index }\label{contributionSI}

In principle, intrinsic spatial variations in the polarized dust spectral index can also contribute to the overall observed dispersion in $\beta$. As discussed in \S{\ref{theory}}, variations in the intrinsic polarized dust spectral index can be attributed to a plethora of different dust microphysics, including, for example, variations in dust composition, grain sizes, and orientation with respect to local radiation/magnetic field geometries. We find that since Planck instrument noise can account for most of the observed dispersion in $\beta$ in \cite{Ade2015b}, the total error budget in $\beta$ does not require a contribution from intrinsic variations in the polarized dust spectral index. 

\section{Summary and Discussion}\label{future}

Our main results are summarized below:

\begin{enumerate}
	\item We showed that multiple line-of-sight contributions from dust clouds of different temperature and polarization angle orientations can lead to significant decorrelation in the observed line-of-sight integrated polarization parameters (i.e. the observed Stokes parameters, polarization fraction and polarization angle) when the polarized dust SED is extrapolated from 350 Ghz to 150 GHz, resulting in \textit{line-of-sight extrapolation noise} (\S{\ref{theory}}).
	\item We performed a Monte Carlo analysis using two different dust distribution models to estimate the statistical properties of the \losen, and found that both models are consistent with each other, producing approximately the same degree of \losen, with 68th percentile errors of $\sim7\%$ in $Q$, $\sim3\%$ in $p$ and $\sim1\degree$ in $\alpha$ when extrapolating dust properties from 350 GHz to 150 GHz. However, the distribution of the \losen\ is non-Gaussian with long tails, implying that sightlines with very large \losen\ are more likely to occur than expected from a Gaussian distribution (\S{\ref{method}} -\S{\ref{results}}, \Rf{corner.pdf}, \Rf{model_combine.pdf}).
	\item We extended the fiducial models to account for variations in the distribution of contributing clouds along the line-of-sight, and quantified the \losen\ in each variation (\S{\ref{fiducialext}}).
	\item We investigated the dependence of \losen\ on the input parameters in our Monte Carlo analysis, and found the \losen\ to be \textit{most sensitive to the temperature distribution} of the dust clouds, and least sensitive to the distribution of polarization fractions of the dust clouds (\S{\ref{senseinput}}).
	\item We explored various potential systematics, including variations in the dust temperature distribution with distance, the choice of the angular resolution of the Pan-STARRS 1 dust reddening map used in model 2, and the effect of Gaussian smoothing and degradation of the angular scale of the pixels in the generated $I$, $Q$ and $U$ maps, and found the statistical properties of the \losen\ to be \textit{insensitive} to these effects \S{\ref{Tsystematics}}-\S{\ref{mapsystematics}}. 
	\item We estimate that the \losen\ is approximately $4\%$ of the total extrapolation uncertainty in the polarized dust power reported by the Planck analysis \cite{PlanckCollaboration2014}, and is not a significant error source compared to current Planck instrument noise \S{\ref{Planckcomparison}}.  
\end{enumerate}
 
Based on our current analysis, the \losen\ is about an order of magnitude smaller than the instrument noise, and about 4\% of the total extrapolation uncertainty reported by the Planck experiment. In this current noise-limited regime, the \losen\ is small compared to instrument sensitivity constraints. However, future CMB experiments like CMB S4 \cite{Abazajian2015} will drastically improve the instrument sensitivity and reduce instrument systematics, allowing us to perhaps enter a regime where \losen\ becomes a significant foreground uncertainty. 

In particular, if the inflationary B-mode signal is comparable to or below \losen\ levels, accounting for this source of uncertainty becomes potentially important. We can estimate the approximate level at which \losen\ becomes comparable to the inflationary B-mode signal by scaling the Planck/BICEP2 results as follows: When extrapolating the dust B-mode power $D^{BB}_{\ell}$ from 353 GHz to 150 GHz in the BICEP2 field, the total dispersion in the observed polarized dust spectral index $\beta$ results in an extrapolation uncertainty of $(+0.28, -0.24)\times10^{-2} \unit{\mu K_{CMB}^2}$ in $D^{BB}_{\ell}$. If \losen\ is $\sim$4\% of the total extrapolation uncertainty, its contribution to the extrapolation uncertainty in $D^{BB}_{\ell}$ is  $(+1.2, -0.99) \times 10^{-4} \unit{\mu K_{CMB}^2}$, or approximately on the order of $\pm 10^{-4} \unit{\mu K_{CMB}^2}$. 

On the other hand, the expected CMB primordial B-mode power at $\ell=80$ for a tensor-to-scalar ratio of $r=1$ is $6.71 \times 10^{-2} \unit{\mu K_{CMB}^2}$} \cite{PlanckCollaboration2014}. The primordial B-mode power scales linearly with $r$, so a CMB primordial B-mode spectrum at $\ell=80$ for $r \approx 0.0015$ would have a power of $10^{-4} \unit{\mu K_{CMB}^2}$, comparable to the noise contribution from the \losen. This implies that in order to achieve a detection of the primordial B-mode signal at scales of $r\lesssim0.0015$, \losen\ becomes a significant source of noise that has to be accounted for. This simple scaling analysis assumes that dust foreground separation uses dust maps at 350 GHz extrapolated down to 150 GHz. In principle, the \losen\ would be larger if the frequency range being extrapolated over increases (for example, down to 95 GHz for BICEP3 \cite{Ahmed2014}).

As previously stated, the \losen\ is part of an astrophysical error floor that cannot be reduced by virtue of better instrumental sensitivity alone. However, there are mitigating strategies that can be used to reduce this astrophysical systematic uncertainty. As discussed in \S{\ref{resfid}}, if the distribution of the \losen\ is non-Gaussian with long tails, one possible strategy is to use information from magnetic field tomography to identify non-predictive sightlines on the tails of that distribution, where the \losen\ is likely to be large. By mapping the polarization of starlight from stars at known distances along a line-of-sight, we can, in principle, reconstruct the magnetic field geometry along that line-of-sight. If such a study is conducted on regions targeted by CMB experiments, we can infer the 3D polarization orientation of dust clouds in that target region. Since the \losen\ tends to be more significant along lines-of-sight where the polarization angle of dust clouds are misaligned with respect to each other (see e.g. \S{\ref*{senseinput}}), one strategy to reduce the \losen\ is to discern regions where the magnetic fields are particularly misaligned along the line-of-sight, and mask out these regions in CMB analyses. Future magnetic field tomography experiments, such as PASIPHAE \cite{Tassis2016} will play an important role in these efforts to ameliorate this source of uncertainty. 

{\it Note added:} After this paper was completed, the Planck collaboration released a study of decorrelation in dust polarization properties between frequencies due to spatial variations in the polarized dust SED \cite{PlanckCollaboration2016a}. The \losen\ discussed here could be responsible for at least part of this decorrelation; therefore the models introduced here complement their analysis. They pointed out that inaccurate extrapolation of polarized dust properties between frequencies can result in a positively biased estimate of the tensor-to-scalar ratio, $r$. This underscores the importance of the effect treated here.

\bigskip

\begin{acknowledgments}
We thank Vasiliki Pavlidou, Konstantinos Tassis, Nikos Kylafis, Brandon Hensley, Jo Dunkley and Ben Thorne for very helpful comments. JP gratefully acknowledges the use of the Seaborn \cite{Waskom2014} and Corner.py \cite{Foreman-Mackey2016} plotting libraries in this work. This work made use of computing resources and support provided by the Research Computing Center at the University of Chicago. The work of SD is supported by the U.S. Department of Energy, including grant DE-FG02-95ER40896. This work was supported in part by the Kavli Institute for Cosmological Physics at the University of Chicago through grant NSF PHY-1125897 and an endowment from the Kavli Foundation and its founder Fred Kavli.
\end{acknowledgments}

\bibliography{reference}

\appendix*
\section{\label{Tpvalues} Fiducial $T$ and $p$ distribution parameters} 
\begin{table*}[b]
	\begin{ruledtabular}
		\begin{tabular}{ c c c c c c c c }
			\multicolumn{4}{c}{Poisson Model (Model 1)} & \multicolumn{4}{c}{Pan-STARRS 1 Dust Reddening Map (Model 2)} \\
			\cline{1-4} \cline{5-8}
			Clouds/kpc & $\mathrm{T_{mean}}$ &  $\mathrm{\sigma_T}$ & $\mathrm{p_{mean}}$ & Cumulative Distance Modulus& $\mathrm{T_{mean}}$ &  $\mathrm{\sigma_T}$ & $\mathrm{p_{mean}}$ \\
			\hline
			1   &19.700	&1.53	&0.066	&4.0	&19.700	&1.40	&0.060 \\
			2	&19.682	&1.75	&0.075	&4.5	&19.695	&1.70	&0.076 \\
			3	&19.664	&2.02	&0.087	&5.0	&19.690	&2.05	&0.093\\
			4	&19.646	&2.22	&0.100	&5.5	&19.685	&2.40	&0.110 \\
			5	&19.628	&2.45	&0.110	&6.0	&19.680	&2.73	&0.130 \\
			6	&19.610	&2.66	&0.121	&6.5	&19.675	&2.94	&0.137 \\
			7	&19.592	&2.87	&0.130	&7.0	&19.670	&3.20	&0.150 \\
			8	&19.574	&3.05	&0.138	&7.5	&19.665	&3.30	&0.155 \\
			9	&19.556	&3.19	&0.146	&8.0	&19.660	&3.40	&0.155 \\
			10	&19.538	&3.35	&0.154	&8.5	&19.655	&3.42	&0.155 \\
			11	&19.520	&3.52	&0.161	&9.0	&19.650	&3.45	&0.155 \\
			12	&19.502	&3.60	&0.167	&9.5	&19.645	&3.48	&0.157 \\ 
			13	&19.484	&3.71	&0.174	&10.0	&19.640	&3.45	&0.157 \\
			14	&19.466	&3.81	&0.179	&10.5	&19.635	&3.48	&0.157 \\ 
			15	&19.448	&4.00	&0.186	&11.0	&19.630	&3.50	&0.157 \\
			16	&19.430	&4.15	&0.193	&11.5	&19.625	&3.60	&0.157 \\
			17	&19.412	&4.22	&0.198	&12.0	&19.620	&3.65	&0.157 \\
			18	&19.394	&4.35	&0.201	&12.5	&19.615	&3.65	&0.157 \\
			19	&19.376	&4.53	&0.206	&13.0	&19.610	&3.65	&0.157 \\
			20	&19.358	&4.64	&0.211	&13.5	&19.605	&3.65	&0.157 \\
			21	&19.340	&4.70	&0.214	&14.0	&19.600	&3.65	&0.157 \\
			22	&19.322	&4.79	&0.217	&14.5	&19.595	&3.65	&0.157 \\
			23	&19.304	&4.95	&0.222	&15.0	&19.590	&3.65	&0.157 \\
			24	&19.286	&5.00	&0.228	&15.5	&19.585	&3.65	&0.157 \\
			25	&19.268	&5.14	&0.233	&16.0	&19.580	&3.65	&0.157 \\
			26	&19.250	&5.30	&0.238	&16.5	&19.575	&3.65	&0.157 \\ 
			27	&19.232	&5.40	&0.240	&17.0	&19.570	&3.65	&0.157 \\
			28	&19.214	&5.48	&0.244	&17.5	&19.565	&3.65	&0.157 \\
			29	&19.196	&5.65	&0.246	&18.0	&19.560	&3.65	&0.157 \\
			30	&19.178	&5.70	&0.248	&18.5	&19.555	&3.65	&0.157 \\ 
			 & & & &19.0	&19.550	&3.65	&0.157 
		\end{tabular}
	\end{ruledtabular}
	\caption{\label{TPtable}Summary of the best-fit Gaussian distributions for temperature $T$ and polarization fraction $p$ for different variations of two dust distribution models. For model 1, $T$ and $p$ are refitted for different distributions of mean cloud number along a line-of-sight. For model 2, $T$ and $p$ are refitted for different cumulative distance moduli bins. For both models, the standard deviation in the polarization fraction remained unchanged from the fiducial model, $\sigma_p=0.03$. Details are discussed in \S{\ref{fiducialext}}. }
\end{table*}

\end{document}